\documentclass[11pt]{article}
\pdfoutput=1

\usepackage{euscript}
\usepackage{amssymb}
\usepackage{amsfonts}
\usepackage{amsbsy}
\usepackage{amsmath}
\usepackage{epsfig}
\usepackage{amsthm}
\usepackage{amscd}
\usepackage{amstext}

\usepackage[pdftex]{hyperref}

\def\ben{\begin{equation}}
\def\een{\end{equation}}
\def\half{{\textstyle{1\over2}}}

\let\w=\omega    
   
\let\C=\Chi

\let\pa=\partial
\def\be{\begin{equation}}
\def\ee{\end{equation}}
\def\ba{\begin{array}}
\def\ea{\end{array}}

\def\dalemb#1#2{{\vbox{\hrule height .#2pt
       \hbox{\vrule width.#2pt height#1pt \kern#1pt
               \vrule width.#2pt}
       \hrule height.#2pt}}}

\newcommand{\bea}{\begin{eqnarray}}
\newcommand{\eea}{\end{eqnarray}}

\newcommand{\Tr}{{\rm Tr} }

\def\R{{{\Bbb R}}}
\def\C{{{\Bbb C}}}

\def\Z{{{\Bbb Z}}}

\def\Lag{{\mathcal{L}}}

\def\ocal{{\mathcal{O}}}

\def\hY{\hat Y}

\def\IC{\C}
\def\ICP{\C{\Bbb{P}}}
\def\IZ{\Z}
\def\CN{{\cal N}}

\begin{document}

\begin{flushright}
arXiv:0901.1160 [hep-th]
\end{flushright}

\begin{center}

\vspace{1cm} { \LARGE {\bf Landscape of superconducting membranes}}

\vspace{1cm}

Frederik Denef$^{\sharp, \flat}$ and Sean A. Hartnoll$^{\sharp}$

\vspace{0.8cm}

{\it ${}^\sharp$ Jefferson Physical Laboratory, Harvard University,
\\
Cambridge, MA 02138, USA \\}
\vspace{0.5cm}

{\it ${}^\flat $ Instituut voor Theoretische Fysica, U Leuven, \\
Celestijnenlaan 200D, B-3001 Leuven, Belgium \\}

\vspace{0.6cm}

{\tt  denef\#physics.harvard.edu}  \hspace{0.5cm} {\tt hartnoll\#physics.harvard.edu} \\

\vspace{2cm}

\end{center}

\begin{abstract}

The AdS/CFT correspondence may connect
the landscape of string vacua and the `atomic landscape' of
condensed matter physics. We study the stability of
a landscape of IR fixed points of
${\mathcal{N}}=2$ large $N$ gauge theories in 2+1 dimensions,
dual to Sasaki-Einstein compactifications of M theory,
towards a superconducting state. By exhibiting instabilities
of charged black holes in these compactifications,
we show that many
of these theories have charged operators that
condense when the theory is placed at a finite chemical potential.
We compute a statistical distribution of critical superconducting temperatures
for a subset of these theories. With a chemical potential of one milliVolt,
we find critical temperatures ranging between $0.24$ and $165$
degrees Kelvin.

\end{abstract}

\pagebreak
\setcounter{page}{1}

\section{A tale of two landscapes}

This paper will explore the relation between quantum critical
phenomena in condensed matter systems and the landscape of string
vacua. The connection between these will be the AdS/CFT
correspondence \cite{Maldacena:1997re}.

String theory has infinitely many compactifications to four dimensions. Of those, googols may
lead to low energy physics compatible with observations \cite{Bousso:2000xa, Kachru:2003aw,
Susskind:2003kw, Douglas:2003um, Douglas:2006es, Denef:2008wq}.
The existence of this landscape of string theory vacua
has lead to a revival of anthropic reasoning in cosmology and particle physics, together with associated
philosophical conundrums and worries about the scientific status
and predictability of string theory. Against this background, it
would be appealing if the string landscape could be related to a
different set of physical systems than particle physics and
cosmology.

Whereas particle physics and cosmology give us direct experimental access to only one vacuum and its associated low energy effective field theory, in condensed matter physics there is a virtually unlimited supply of `vacua' and corresponding field theories. Typical examples are crystal lattices. These are metastable ground states of a single underlying microscopic theory, the Standard Model, translation invariant at large distance scales and with low energy excitations described by effective field theories. Material science is in essence the exploration of this vast landscape. In addition, an increasing range of lattice Hamiltonians can be engineered and controlled in tabletop experiments, for instance using optical lattices \cite{Greiner}.

While the systems arising in the `atomic landscape'
are generally sensitive to their underlying discreteness,
as a function of couplings they can undergo second order
phase transitions at zero temperature, called quantum phase
transitions. At the quantum critical point the long distance physics is sometimes
described by a continuum `relativistic' conformal field theory
(CFT), e.g. \cite{sachdev, sachdev2}. We will focus on such
relativistic quantum critical theories
as they are the cases in which AdS/CFT is best understood. Note however
that the AdS/CFT correspondence can be adapted to non-conformal
relativistic theories (see e.g. \cite{Aharony:2002up} for a review) and also to theories
with a non-relativistic scale invariance \cite{Son:2008ye, Balasubramanian:2008dm,
Kachru:2008yh}. We will furthermore focus in this paper on 2+1 dimensional systems.

The AdS/CFT correspondence \cite{Maldacena:1997re, Witten:1998qj,
Gubser:1998bc} implies the existence of a $2+1$ dimensional
conformal field theory for every $3+1$ dimensional theory of
quantum gravity in an asymptotically Anti-de Sitter spacetime. The
string landscape provides an immense number of such theories.
Therefore, the string landscape also provides a wealth of new
quantum critical, that is, scale invariant, theories. Whether any of
these theories can be used to model the physics associated to quantum phase
transitions in experimentally realisable discrete systems is an important question for future work.
In this paper we initiate a study of their properties.

Given a vacuum of a theory, two immediate questions are firstly
to characterise low energy excitations about the vacuum and secondly
to enquire about the stability of the vacuum configuration. These two
issues can be directly related. For instance, in conventional superconductivity
an instability of the vacuum with unbroken gauge symmetry arises due to interactions
between low energy phonons and (dressed) electrons.

For generic lattice structures there is by now a very well developed set of
techniques for identifying the low lying degrees of freedom and their dynamics.
Some examples are shown in table 1. However, at quantum critical points
the system is not describable in terms of conventional quasiparticle
degrees of freedom. The critical point
describes the dynamics of highly nonlocal entangled states of
matter, in which different competing orders are finely balanced
\cite{sachdev2}. There is no preferred energy scale and generically no weak
coupling. The lesson of the AdS/CFT
correspondence is that, at least in a `large $N$' limit, there can be a dual
semiclassical description of quantum critical physics\footnote{It is important
to emphasise that unlike in the large $N$ limit of, for instance, the $O(N)$ model,
the AdS/CFT theories are always strongly coupled in the gravity regime.}. Examples of dual
low energy excitations are also shown in table 1.

\begin{table}[h]
\begin{tabular}{| l | l | l |}
\hline & Atomic Landscape & String Landscape \\
\hline \hline Microscopic theory & Standard Model & M theory  \\
\hline Fundamental excitations & Leptons, quarks, photons, etc. & ?? \\
\hline Typical vacuum & Atomic lattice & Compactification \\
\hline Low energy excitations & Dressed electrons, phonons, & Gravitons, gauge bosons, \\
& spinons, triplons, etc. & moduli, intersectons, etc.  \\
\hline
Low energy theory & Various QFTs & Various supergravities \\
\hline
\end{tabular}
\caption{Comparison of two landscapes.}
\end{table}

Table 1 suggests a complementary relationship between the
string and atomic landscapes. The string landscape may supply tractable
models of quantum critical points in the atomic landscape.
Furthermore, studying the string landscape in its totality 
may lead to the identification of universal or typical properties and also novel
exotic behaviors. One is
also lead to wonder whether the atomic landscape might have implications
for string theory. We will speculate on this latter connection at the end of the paper.

The dynamical property of CFTs with string vacuum duals that we shall investigate in this paper
is the potential instability towards a superconducting phase. We show that
a large class of string compactifications do indeed have such instabilities.
Inter alia these backgrounds provide explicit string theory realisations of
holographic s wave superconductors \cite{Gubser:2008px, Hartnoll:2008vx,
Hartnoll:2008kx}, including cases in which the dual field theories are
known.\footnote{Making approximations to the nonabelian DBI action, holographic p wave
superconductors \cite{Gubser:2008zu, Gubser:2008wv, Roberts:2008ns} can be obtained in string
theory using coincident D branes \cite{Roberts:2008ns, Ammon:2008fc, Basu:2008bh}.} In particular, the theories are those arising on M2 branes
placed at the tip of a Calabi-Yau cone. These are the IR fixed
points of ${\mathcal{N}}=2$ supersymmetric gauge theories in 2+1 dimensions.
Among these, we find a superconducting instability in the maximally
supersymmetry ${\mathcal{N}}=8$ CFT in 2+1 dimensions
at a finite chemical potential.

We will begin by reviewing the framework of holographic
superconductivity. We then go on to discuss a subset of the landscape
given by ${\mathcal{N}}=2$ Freund-Rubin Sasaki-Einstein compactifications of M theory. These theories can be consistently truncated to Einstein-Maxwell theory on a four dimensional space with negative cosmological constant.
We show that there exist minimally coupled charged pseudoscalar modes that decouple from
all other fluctuation modes at the linearized level, in arbitrary backgrounds solving the Einstein-Maxwell equations. They correspond to modes of the M theory 3-form obtained by reducing certain harmonic 4-forms on the Calabi-Yau cone over the Sasaki-Einstein manifold. We show that these modes lead to instabilities towards a superconducting phase of the dual CFT at low temperatures for a large number of Sasaki-Einstein compactifications, and we obtain a distribution of critical temperatures on this landscape.

\section{Holographic superconductors}

\subsection{General framework}

Holographic superconductors are a class of quantum critical theories which have
an instability to a superconducting phase at low temperatures when
held at a finite chemical potential $\mu$ \cite{Gubser:2008px, Hartnoll:2008vx,
Hartnoll:2008kx}. One can equivalently work with a fixed charge density $\rho$.
Scale invariance and dimensional
analysis imply that the critical temperature $T_c \propto \mu$. Our objective
is to show that a large number of simple string vacua are holographic superconductors
and to determine $T_c/\mu$ for these theories.

The minimal bulk action for a holographic superconductor must
describe the dynamics of the metric, a Maxwell field and at least one charged
field that can condense and spontaneously break the $U(1)$ symmetry.
We focus in this work on the case in which the charged field is a scalar
in $AdS_4$. In general the full nonlinear action is complicated, as a consistent embedding
into string theory will typically involve many coupled fields. Physically
this implies that there will be many condensates at low temperature. In this work we
avoid this problem by only considering the scalar equations of motion to linearised order,
at which many fields decouple. This is sufficient to determine the critical temperature.

The bulk action for a minimally coupled scalar field to quadratic order in the scalar is
\be\label{eq:action}
\Lag = \frac{M^2}{2} R + \frac{3 M^2}{L^2} - \frac{1}{4g^2} F_{\mu\nu} F^{\mu\nu}
- \left|\nabla \phi - i q A \phi \right|^2 - m^2 \left| \phi \right|^2 \,.
\ee
There are four dimensionless quantities in this action: the AdS radius in Planck units
$(ML)^2$, the mass squared of the scalar field $(mL)^2$, the Maxwell coupling $g$
and the charge of the scalar field $q$. We will show in the following section that this
action can be consistently obtained from M theory Freund-Rubin compactifications. The
internal geometry of the compactification will fix the values of these coefficients. The
dimensionless quantities have the following field theory interpretations:
\begin{itemize}

\item  The central charge of the CFT is
\be\label{eq:central}
c = 192 (ML)^2 \,, \quad \text{where} \quad s = \frac{c \pi^3}{54} T^2 \,.
\ee
Here $s$ is the entropy density. Recall that for a 2+1 CFT, the central charge can be defined in two ways \cite{Cardy:1987dg}.
Either as a parametrisation of the energy momentum tensor two point function, or
as a parametrisation of the entropy density, as we have used in (\ref{eq:central}). It was noted in
\cite{Kovtun:2008kw} that these two notions agree for theories with classical
gravity duals.\footnote{In equation (\ref{eq:central}) we are using the normalisation of \cite{Kovtun:2008kw}  for the central charge. In this normalisation, the central charge of a massless free boson is $c=81 \zeta(3)/\pi^4 \approx 0.9996$.}

\item The electrical conductivity at zero momentum is frequency independent \cite{Herzog:2007ij}
\be
\sigma \equiv \sigma_{xx} = \frac{1}{g^2} \,.
\ee
This is the conductivity appearing in Ohm's law $j = \sigma E$. Recall that conductivity is dimensionless in 2+1 dimensions, and so $\sigma$ may also be thought of as a central charge.

\item The scaling dimension of the charged operator $\ocal$ dual to the bulk field $\phi$ is \cite{Witten:1998qj,Gubser:1998bc}
\be
\Delta (\Delta - 3) = (mL)^2 \,.
\ee
Both roots to this equation are admissible \cite{Klebanov:1999tb} so long as they satisfy the unitarity bound $\Delta \geq \frac{1}{2}$.

\item The charge $q$ is the charge of the dual operator $\ocal$. We will consider cases in which the gauge group is $U(1)$ (rather than $\R$) and work in units in which the charges take integer values.

\end{itemize}

The quantum critical theory at finite temperature and chemical potential
is dual to the bulk theory in an AdS-Reissner-Nordstom black hole background.
This has metric
\be\label{eq:RNblackhole}
ds^2 = - f dt^2 + \frac{dr^2}{f} + \frac{r^2}{L^2}(dx^2+ dy^2) \,,
\ee
and scalar potential
\be
A_0 = \mu \left(1 - \frac{r_+}{r}\right) \,.
\ee
The function $f$ is given by
\be
f = \frac{r^2}{L^2} - \left(\frac{r_+^2}{L^2} + \frac{\mu^2}{2 g^2 M^2} \right)
\frac{r_+}{r} + \frac{\mu^2}{2 g^2 M^2} \frac{r_+^2}{r^2} \,,
\ee
where the horizon radius $r_+$ is related to the temperature through
\be
T = \frac{1}{8 \pi r_+} \left(\frac{6 r_+^2}{L^2} - \frac{\mu^2}{g^2 M^2} \right) \,.
\ee
Here $T$ and $\mu$ are the temperature and chemical potential of the field theory, respectively. The charge density of the field theory is
\begin{equation}
 \rho = \frac{\mu r_+}{g^2 L^2} = \mu \sigma T \left( \frac{2 \pi}{3} + \sqrt{\left( \frac{2 \pi}{3} \right)^2 + \frac{32 \sigma}{c} \frac{\mu^2}{T^2} } \, \right) \, .
\end{equation}
To see whether the theory develops superconductivity we need to check the stability of this
background against fluctuations of the scalar field.

\subsection{Criterion for instability of minimally coupled scalars}
\label{sec:stability}

The equations of motion for the charged scalar field following from (\ref{eq:action}) are
\be
- \left(\nabla_\mu - i q A_\mu \right) \left(\nabla^\mu - i q A^\mu
\right)\phi + m^2 \phi = 0 \,.
\ee
Looking for an unstable mode of the form $\phi = \phi(r) e^{- i \w t}$ one obtains
\be\label{eq:scalar}
- \phi'' - \left( \frac{2}{r} + \frac{f'}{f} \right) \phi' - \frac{[r \w + q \mu (r-r_+)]^2}{r^2 f^2} \phi + \frac{m^2}{f} \phi
= 0 \,.
\ee
The AdS-Reissner-Nordstrom black hole will be unstable if there is a normalisable solution to this
equation, with ingoing boundary conditions at the horizon, such that $\w$ has a nonzero positive imaginary part.

We will shortly solve (\ref{eq:scalar}) numerically. A few prior comments are in order. It is useful to introduce the ratio\footnote{Essentially this ratio was also considered in \cite{Kovtun:2008kx}.}
\be\label{eq:ratio}
\gamma^2 \equiv \frac{c}{96 \sigma} = 2 g^2 (ML)^2 \,.
\ee
A ratio of central charges, $\gamma$ might be thought of as quantifying the efficiency of charge transport
in the theory. The BPS bound\footnote{This bound can be derived from the superconformal algebra when the $U(1)$ under consideration is the R-symmetry in this algebra, as will in fact be the case for the Sasaki-Einstein compactifications we will consider. Unlike in asymptotically flat space, the BPS bound lies strictly below the black hole extremality bound \cite{Romans:1991nq}, except in the limit $q \to 0$. Extremal black holes do not preserve any supersymmetry.} for charged scalars can then be written as
\be
\Delta \geq \gamma q \,.
\ee
The normalisation can be obtained, for instance, from the extremality condition of black holes
with spherical horizons that are much smaller than the AdS radius in the theory (\ref{eq:action}).
Recall that $q$ is quantised to be integer. We further observe, allowing ourselves to rescale the radial coordinate, that the equation (\ref{eq:scalar}) depends only on the following three dimensionless quantities: $\Delta$, $\gamma q$ and $\gamma T / \mu$. Fixing the first two of these, the mass and charge, we solve (\ref{eq:scalar}) to obtain the critical temperature $T_c$ below which there is an instability. In more detail, the numerical algorithm proceeds as follows. We fix $\gamma T/\mu$, $\gamma q$ and $\Delta$, and start by constructing the solution in the very near horizon region obtained by Taylor series expansion to third order in the coordinate distance from the horizon. We then numerically solve the linear differential equation (\ref{eq:scalar}) out to a sufficiently large value of $r$. 
The equation is solved with $\w=0$ as we are looking for the onset of an instability.
Finally, we match this to the general large $r$ asymptotic solution, obtained by power series expansion to seventh order. (Working to such high order is necessary to get accurate results across the full parameter range.) This procedure thus yields two coefficients as a function of $T$, multiplying the solutions with $r^{-\Delta}$ and $r^{\Delta-3}$ leading asymptotics.  Solving for the largest value of $T$ for which the coefficient multiplying the $r^{\Delta-3}$ branch vanishes gives us $T_c$ at the given values of $\Delta$ and $\gamma q$. This is then repeated for a fine grid of values of $\Delta$ and $\gamma q$. The result is shown in figure 1.

\begin{figure}[h]

\begin{center}
\includegraphics[height=10cm]{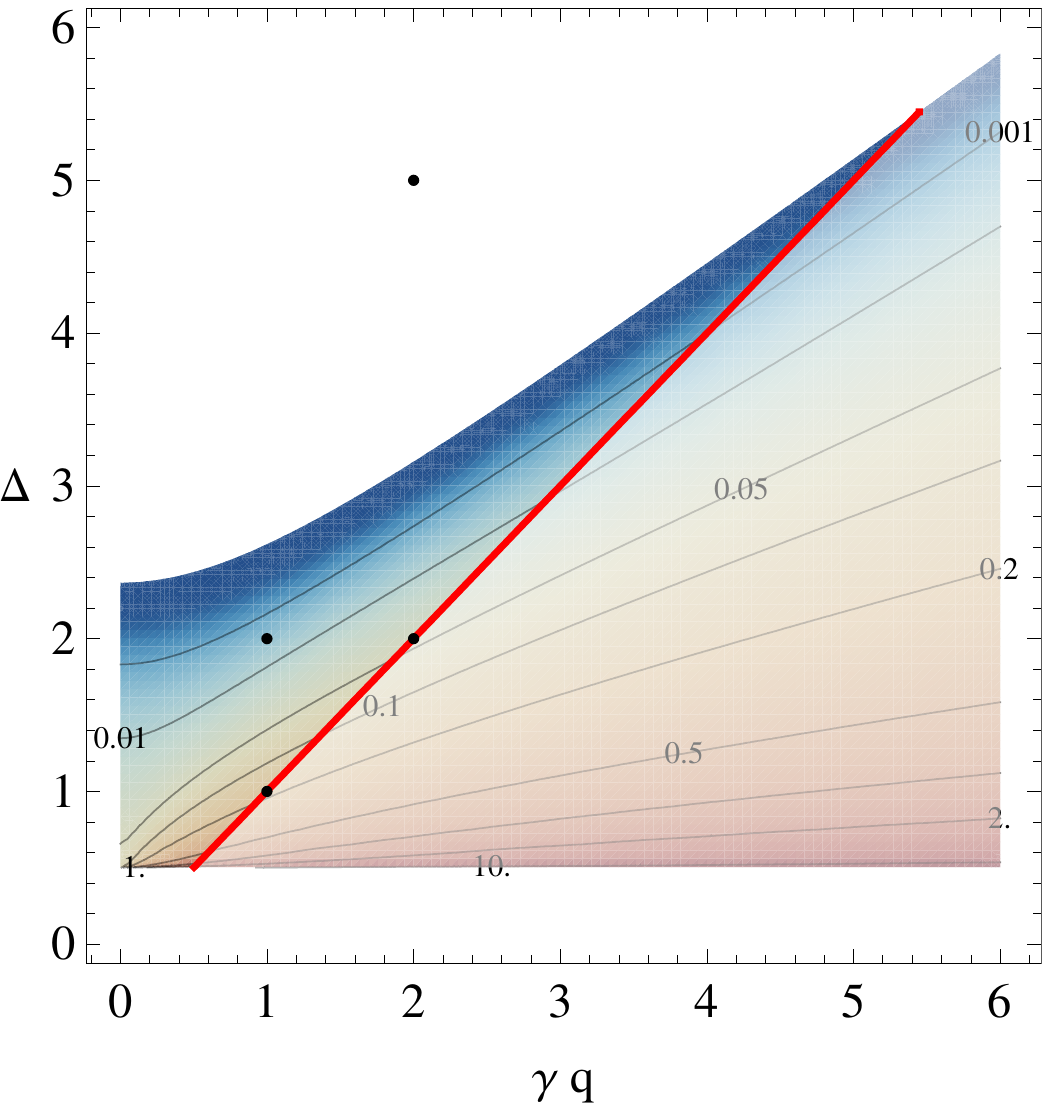}
\end{center}
\caption{The critical temperature $T_c$ for a minimally coupled scalar as a function of the charge $\gamma q$ and dimension $\Delta$ of the dual operator. Contours are labeled by values of $\gamma T_c /\mu$. The BPS line $\Delta = \gamma q$ is shown in red; the shaded triangle to the left of it is the window of unstable values compatible with the BPS bound. The top boundary $q^2 \gamma^2 = 3 + 2 \Delta (\Delta-3)$ is a line of quantum critical points separating superconducting and normal phases at $T=0$. The bottom boundary is the unitarity bound $\Delta=1/2$, where $T_c$ diverges. The black dots indicate special cases which we will see arise in the context of ${\mathcal{N}}=2$ M2
brane theories. \label{fig:map}}
\end{figure}

The zero temperature result of this plot can be understood analytically. If we look for a threshold unstable mode, with $\w=0$, at zero temperature, then near the horizon we find the behaviour
\be\label{eq:nearhorizon}
\phi \sim (r-r_+)^{(-3 \pm \sqrt{3} \sqrt{3 - q^2 \gamma^2 + 2 \Delta (\Delta - 3)})/6} \,.
\ee
On general grounds one expects an instability to arise when the field oscillates infinitely many times before reaching the horizon \cite{Gibbons:2002pq}. From (\ref{eq:nearhorizon}) we see that this requires
\be\label{eq:criterion}
q^2 \gamma^2 \geq 3 + 2 \Delta (\Delta-3) \,.
\ee
Therefore we expect an instability when the charge of the scalar field is sufficiently large as given by (\ref{eq:criterion}). If the charge is lower than the critical value there will never be an instability, as raising the temperature acts to stabilise the theory. The black line in figure 1, obtained numerically, is precisely the curve (\ref{eq:criterion}) separating stable backgrounds from backgrounds that become unstable below some temperature. This is a line of quantum critical points. It would be interesting to study in detail the dynamics close to these points.

The instability criterion (\ref{eq:criterion}) reduces to the inequality noted in \cite{Hartnoll:2008kx} for the case of neutral scalar fields ($q=0$). There the result was obtained by comparing the mass squared of the field to the Breitenlohner-Freedman bound in the $AdS_2$ near horizon region. The full result (\ref{eq:criterion}) may be obtained by requiring the near horizon effective mass squared, including the coupling to the Maxwell field \cite{Gubser:2008pf}, to be below the $AdS_2$ Breitenlohner-Freedman bound.

\begin{figure}[h]
\begin{center}
\includegraphics[height=6cm]{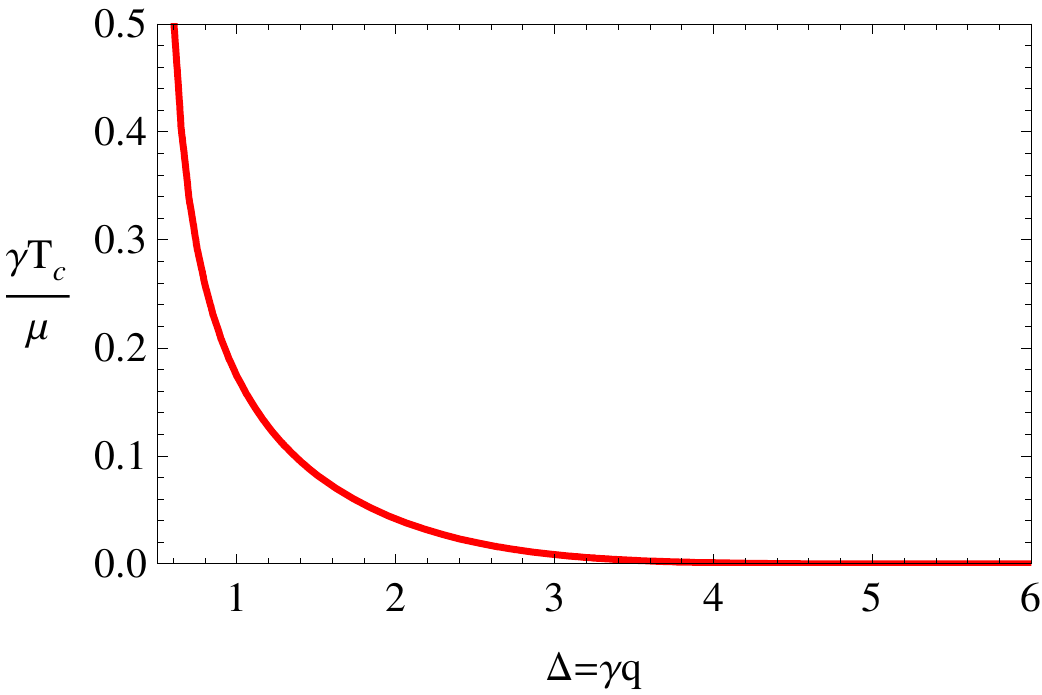}
\end{center}
\caption{Critical temperature $\gamma T_c/\mu$ as a function of $\Delta$ for operators on the BPS line $\Delta=\gamma q$. \label{fig:BPSline}}
\end{figure}

The remaining noteworthy feature of figure \ref{fig:map}, is that the critical temperature diverges as $\Delta \to \half$. This divergence is exhibited clearly in figure \ref{fig:BPSline}, which shows
the critical temperature as a function of operator dimension along the BPS line $\Delta = \gamma q$. It is presumably related to the fact that $\Delta  = \half$ modes form singleton representations of the AdS$_4$ isometry group. These modes can be gauged to the boundary of AdS, which one thinks of as the UV of the field theory, and hence are not sensitive to the temperature, which only affects the IR physics. Thus the superconducting instability can never be stabilised by the temperature in this case. The field theory statement of this fact is that these modes are decoupled from all others and therefore do not acquire a thermal mass.

If we wish to find string theory realisations of holographic superconductivity, we need to find compactifications of string theory that have charged scalars with masses and charges that fall inside the shaded region to the left of the BPS line in figure \ref{fig:map}.

\subsection{The weak gravity bound}
\label{sec:weak}

A priori it is not obvious that there exist compactifications with charged scalars that lie in the left hand region of figure 1. An argument in favour of the generic presence of an instability comes from the conjectured `weak gravity' bound \cite{ArkaniHamed:2006dz}. Perhaps the sharpest of the statements in that paper was the requirement that extremal black holes should be able to decay in consistent theories of quantum gravity. In asymptotically Minkowski spacetime, a simple kinematic argument shows that this requirement implies that there must exist a charged particle in the theory that has mass and charge related by $m \leq \sqrt{2} g q M$, where $m, g, q$ and $M$ have the same meanings as they did in the previous subsections. The interest of this statement is that if $g q \ll 1$, then the charged particle is much lighter than would be predicted from standard effective field theory logic.

In asymptotically Anti-de Sitter spacetimes it is less straightforward to make kinematic arguments for a weak gravity bound, as particles may not scatter out to infinity. However, the criterion (\ref{eq:criterion}) for a classical instability was obtained from only the near horizon geometry of the extremal black hole. If the preferred decay mode of the black hole is through a minimally coupled scalar, as we have been assuming, then (\ref{eq:criterion}) is a natural candidate for the correct weak gravity bound. Namely, in any consistent asymptotically AdS theory of quantum gravity there should exist a charged particle with charge $q$ and energy $\Delta$ such that (\ref{eq:criterion}) is satisfied and extremal black holes can decay. We can note that (\ref{eq:criterion}) does not reduce to the Minkowski space bound when $\Delta \gg 1$. We are only considering large AdS black holes, smaller black holes can require a more
stringent condition in order to decay.

A caveat to the above statement is the possibility of decaying through charged modes that are not minimally coupled scalars. Given a field with a specified spin and coupling to the Maxwell field, it is easy to rerun the above argument involving the near horizon Breitenlohner-Freedman bound and obtain an instability criterion analogous to (\ref{eq:criterion}). The weak gravity bound would only require the existence of one unstable mode, of any spin and coupling.

The instability we are describing is essentially Schwinger pair production. Although this is initially a quantum mechanical effect, once there is sufficient condensate accumulated it is described as a classical instability in terms of macroscopic fields. Whatever the microscopic mechanism for emission of charge from the black hole, it seems likely that the classical field instability considered here is the correct description once the number of quanta involved becomes large.
Furthermore, numerical investigations in \cite{Hartnoll:2008kx} suggested (but not conclusively) that at the endpoint of the extremal black hole instability, if the charge of the scalar field is nonzero, all of the charge is carried by the scalar field condensate. Therefore this instability leads to the complete decay of the extremal black hole, as required by the weak gravity conjecture.

An interesting exception to the statements in the previous paragraph might arise if the preferred decay mode of the black hole were to charged fermionic particles. In the absence of a pairing mechanism these will not develop macroscopic occupation numbers, but rather build up a fermi surface.
This could lead to novel black holes with charged fermionic hair.

Whether or not one believes in the weak gravity bound, we shall now show that there indeed exist a large set of vacua in which extremal AdS-Reissner-Nordstrom black holes are unstable. Note that extremal AdS-Reissner-Nordstrom black holes are not supersymmetric and do not saturate the BPS bound.

\section{Charged scalars from Sasaki-Einstein vacua}

\subsection{${\mathcal{N}}=2$ Freund-Rubin compactifications of M theory}
\label{sec:freund}

The M theory bosonic action is (in the conventions of \cite{Duff:1986hr})
\be
S = \frac{1}{2 \kappa^2} \int d^{11}x \sqrt{-g} R - \frac{1}{\kappa^2} \int \left[ G \wedge \star G +
\frac{2}{3} C \wedge G \wedge G \right] \,,
\ee
with $G = d C$.
We are interested in Freund-Rubin vacua with a background electromagnetic field
in four dimensions. See for instance \cite{Gauntlett:2007ma, Herzog:2007ij}. The metric ansatz is
\be\label{eq:ansatz}
ds^2_{11} = L^2 ds^2_{M_4} + 4 L^2 \left[ (a [d\psi + A] + \sigma)^2 + ds^2_{M_6}\right] \,,
\ee
supported by the flux
\be\label{eq:ansatz2}
G = \frac{3 L^3}{2} \text{vol}_{M_4} - 4 L^3 a \, \w \wedge \star_4 F  \,,
\ee
where $\frac{1}{2} d\sigma=\w $
is the K\"ahler form on $M_6$, which is taken to be
a six real dimensional K\"ahler-Einstein manifold satisfying ${\rm Ric}_{M_6} = 8 \, g_{M_6}$, and $F=dA$. In (\ref{eq:ansatz})
the coefficient $a$ is such that $\psi$ has range $2\pi$, and we have chosen
the four dimensional gauge connection $A$ to be normalised so that excitations have integer charges.

One can check that (\ref{eq:ansatz}) and (\ref{eq:ansatz2}) solve the eleven dimensional
equations of motion if and only if the four dimensional metric $g^{(4)}$ and gauge field $A$
solve the four dimensional Einstein-Maxwell-AdS equations of motion. These come from the
effective four dimensional Lagrangian density
\be\label{eq:4daction}
\Lag^{(4)} = \frac{1}{2 \kappa^2_4} \left[R^{(4)} + \frac{6}{L^2} - 4 L^2 a^2 F_{\mu\nu} F^{\mu\nu} \right] \,,
\ee
where
\be
\frac{1}{2 \kappa^2_4} = \frac{(2L)^7 \text{Vol}(M_7) }{2 \kappa^2} \,.
\ee
In this expression $M_7$ refers to the Sasaki-Einstein manifold\footnote{We will only consider quasi regular Sasaki-Einstein manifolds, i.e.\ those for which the orbits of the Killing vector close. Hence the fibration is $U(1)$ rather than $\R$, and charges are quantised.}
\be\label{eq:M7}
ds^2_{M_7} =  (a d\psi+ \sigma)^2 + ds^2_{M_6} \,,
\ee
with unit radius, that is, $ds^2_{M_7}$ is such that the cone
\begin{equation} \label{M8cone}
 ds_{M_8}^2 = dr^2 + r^2 ds^2_{M_7} \,,
\end{equation}
is Ricci flat, i.e.\ a Calabi-Yau fourfold. The construction of Sasaki-Einstein manifolds as $U(1)$ fibrations
over K\"ahler-Einstein manifolds is reviewed with differing emphases in \cite{Gibbons:2002th,
Boyer:2004fc,Boyer:2008vf,Martelli:2006yb}. The simplest example is $M_7 = S^7$, the round 7-sphere, for which $M_8 = \IC^4$, $M_6 = {\mathbb{CP}}^3$, $ds^2_{M_6}$ the Fubini-Study metric, and $a=1$.
The $U(1)$ symmetry is the R-symmetry of the dual ${\mathcal{N}}=2$ field theory.

In checking that this ansatz indeed provides a consistent truncation to four dimensional Einstein-Maxwell with
a negative cosmological constant, it is important to be precise about
orientations. We are taking the Sasaki-Einstein metric to be orientated such
that its volume form is
\be
\text{vol}_{M_7} =  + \frac{a}{6}  d\psi \wedge \w \wedge \w \wedge \w \,.
\ee
This implies, for instance, that
\be
\star_7 (\w \wedge \w) = + 2 (a d\psi + \sigma) \wedge \w \,,
\ee
which is an equation one uses in confirming consistency.

Comparing the effective action (\ref{eq:4daction}) to our general expression in section
2 above we find that for these theories the `ratio of central charges'
\be\label{eq:gtoa}
\gamma = \sqrt{\frac{c}{96 \sigma}} = \frac{1}{2a} \,.
\ee
The coefficient $\gamma$ is therefore determined by a single component
of the Sasaki-Einstein metric, giving the (constant) radius of the canonical $U(1)$
fibration. This radius is determined topologically. Concretely:
\be\label{eq:gammatop}
\gamma = \frac{2 k}{\gcd c_1(M_6)} \,,
\ee
where $k$ is a positive integer and $\gcd c_1(M_6)$ is the greatest integer by which the first Chern class $c_1(M_6)$ can be divided such that it remains an integral (orbifold) cohomology class \cite{Martelli:2006yb}. The freedom to choose $k$ corresponds to the freedom to quotient the circle by $\IZ_k$. For example for $M_7 = S^7/\IZ_k$, since $\gcd c_1(\ICP^3) = 4$, we get $\gamma=k/2$.
There is a constraint on the values of $k$ that are compatible with supersymmetry. The Killing spinor has
a $\psi$ dependence of the form $e^{i 2 a \psi}$ \cite{Gibbons:2002th}. In order for the spinor to be well defined we must therefore have $4 a \in \Z$. This constrains $k$ not to be too large, given $\gcd c_1(M_6)$. 

By comparison with section 2 we can also obtain the central charge
\be\label{eq:centralcharge}
c = \frac{192 \, L^2}{\kappa^2_4} = \frac{32 \, \pi}{\sqrt{6} \text{Vol}(M_7)^{1/2}} \, N^{3/2} \,.
\ee
In this expression we introduced the M2 brane charge $N \propto \int [*G + C \wedge G]$, which is a positive integer\footnote{Specifically, $N = 3 (2 L)^6 \text{Vol}(M_7)/(2\pi^3 \kappa^4)^{1/3}$. This normalisation can be obtained from the Dirac quantisation condition
for M2 and M5 branes in M theory.}. The dual 2+1 dimensional CFT, to be discussed below, will have an ultraviolet description
as a gauge theory with an $SU(N)$ gauge group.
Like the fiber radius $a$, the normalized volume ${\rm Vol}(M_7)$ can be computed topologically \cite{Martelli:2006yb}.
Bishop's theorem implies that $\text{Vol}(M_7) \leq \text{Vol}(S^7) = \pi^4/3$. Therefore the central
charge (\ref{eq:centralcharge}) is always larger than the central charge of the maximally supersymmetric theory, $c_{{\mathcal{N}}=8} \approx 7.2 \, N^{3/2}$.

\subsection{Examples of Sasaki-Einstein manifolds: Brieskorn-Pham links} \label{sec:BPlinks}

A rich landscape of examples of Sasaki-Einstein manifolds is provided by links of Calabi-Yau hypersurface singularities. These are constructed as follows. Consider a weighted homogeneous polynomial $F(z)$ in $\C^5$. That is, satisfying
\be\label{eq:weighted}
 F(\lambda^{w_1} z_1, \ldots, \lambda^{w_5} z_5) = \lambda^d F(z_1,
\ldots, z_5) \,,
\ee
where $w_i$ and $d$ are positive integers. An example is
\begin{equation} \label{example}
 F(z) = z_1^2 + z_2^5 + z_3^6 + z_4^7 + z_5^8 = 0 \, ,
\end{equation}
which has ${\bf w} = (420, 168, 140, 120, 105)$ and $d=840$. The scaling action implies that the zero set $F(z) = 0$ is a four complex dimensional cone in $\C^5$. By definition, if the hypersurface supports a conical Ricci flat K\"ahler metric as in (\ref{M8cone}), the base (link) of the cone is Sasaki-Einstein.  The $U(1)$ acting as $\psi \to \psi + \Delta \psi$ on (\ref{eq:M7}) acts as $z_i \to e^{i w_i \Delta \psi} z_i$ on the coordinates $z_i$. Thus the integrally quantised charge of the coordinate $z_i$ is precisely $w_i$.
This will shortly enable us to obtain the integrally quantized charge $q$ of various 3-form modes from the weights $\{w_i\}$.

For these Sasaki-Einstein spaces the quantities $a$ and ${\rm Vol}(M_7)$ introduced above, and therefore $\gamma$ and $c$, are known explicitly \cite{Bergman:2001qi,Martelli:2006yb}:
\bea
a & = & \frac{\sum_i w_i - d}{4} \,, \label{awd} \\
\text{Vol}(M_7) & = & \frac{\pi^4 a^4 d}{3 \prod_i w_i} \label{volX7}
\,.
\eea

Not every cone constructed in this manner supports a Ricci flat K\"ahler metric, and correspondingly not every link supports a Sasaki-Einstein metric.
A necessary condition for existence is \cite{Gauntlett:2006vf} $\min_i w_i \geq a>0$, with $a$ given by (\ref{awd}). The CFT interpretation of this bound is quite pretty \cite{Gauntlett:2006vf}: it is the unitarity bound $\Delta \geq \half$ for chiral primaries corresponding to holomorphic functions on the cone. An example that violates this condition is the $A_k$ singularity $F(z)=z_1^{k+1} + z_2^2 + z_3^2 + z_4^2 + z_5^2 = 0$ for $k>2$. A second necessary condition is the bound following
from Bishop's theorem \cite{Gauntlett:2006vf} $\text{Vol}(M_7) \leq \text{Vol}(S^7) = \pi^4/3$, with $\text{Vol}(M_7)$ given by (\ref{volX7}).

A sufficient condition can be formulated \cite{Boyer:2003pe} for the special case of Brieskorn-Pham cones, defined by Fermat type polynomials
\be\label{eq:fermat}
F(z) = z_1^{m_1} + \cdots + z_5^{m_5} = 0 \,.
\ee
These are weighted homogeneous polynomials as in (\ref{eq:weighted}) above with
\be
d = \text{lcm}(m_i | i = 1..5) \,, \qquad w_i = \frac{d}{m_i} \,.
\ee
According to \cite{Boyer:2003pe}, if the coefficients satisfy the
following two conditions, then the link is Sasaki-Einstein:
\be\label{eq:condition}
1 < \sum_i \frac{1}{m_i} < 1 + \frac{4}{3} \min_{i,j}\left\{
\frac{1}{m_i}, \frac{1}{b_i b_j} \right\} \,.
\ee
In this expression
\be
b_j = \gcd(m_j,c_j) \,, \qquad c_j = \text{lcm}(m_i | i \neq j)
\,.
\ee
Furthermore, two such Sasaki-Einstein manifolds, corresponding to
different exponents $\{m_i\}$ and $\{m_i'\}$, are isomorphic if
and only if the two sets of exponents are permutations of each other.
These conditions are sufficient but not necessary for existence. A general
necessary and sufficient condition is not known.

The example given in (\ref{example}) satisfies the conditions (\ref{eq:condition}). It yields a Sasaki-Einstein manifold with $a=28.25$ and ${\rm Vol}(M_7) \approx 0.1396$, so $\gamma \approx 0.0177$ and $c \approx 110 \, N^{3/2}$. The results reviewed in \cite{Boyer:2003qy} imply that this manifold is homotopy equivalent (and therefore, by the generalized Poincar\'e conjecture, homeomorphic) but not diffeomorphic to the standard sphere $S^7$. That is, it can be continuously deformed into the round $S^7$ and there is a continuous but no smooth one to one map to the round $S^7$.
Some further remarkable results about Sasaki-Einstein spaces constructed in this way may be found
in \cite{Boyer:2003ch, Boyer:2003qy, Boyer:2003pe}.

\subsection{Minimally coupled pseudoscalars from 3-form modes}

Given the eleven dimensional background (\ref{eq:ansatz}) and (\ref{eq:ansatz2}) of the previous subsection, we wish to know whether the AdS-Reissner-Nordstrom black hole (\ref{eq:RNblackhole})
is unstable against charged excitations of the background. To answer this question systematically one should consider the general linearised perturbation of the eleven dimensional metric and 3-form about the background. While the spectrum of perturbations about neutral Freund-Rubin compactifications is a well-developed subject \cite{Duff:1986hr}, the analysis is substantially
complicated by the presence of a background four dimensional Maxwell field.
Generically the various modes that appear diagonally in the spectrum about the neutral vacuum are not minimally coupled to the background Maxwell field and furthermore get mixed
amongst each other. For example, one may get couplings such as
$|\phi|^2 F^2$, $\phi \, F^{\mu \nu} \partial_\mu v_\nu$ or $\phi \, \epsilon^{\mu\nu\rho\sigma} F_{\mu \nu} b_{\rho\sigma}$, where $v_\mu$ is some charged vector mode and $b_{\mu\nu}$ a charged 2-form mode. Moreover, such non-minimal couplings tend to qualitatively alter the stability analysis of section \ref{sec:stability}.

Rather then perform the full stability analysis we shall focus on particular 3-form modes which, remarkably, turn out to be only minimally coupled to the Maxwell field and decouple from all other perturbations at the linearised level, in any background satisfying the Einstein-Maxwell equations. We will then show that these modes are sufficient to establish instabilities in a large number of Sasaki-Einstein vacua at finite chemical potential. It should be borne in mind however that there may be more unstable modes than the ones we find. Therefore our results for critical temperatures should be taken as lower bounds only.

In order to describe these modes, it is useful to start with the eight dimensional Calabi-Yau cone (\ref{M8cone}).
Consider a closed self-dual or anti-self-dual 4-form $\hY_4$ on the cone. That is
\be\label{eq:conditions1}
d \hY_4 = 0 \,, \qquad \star_8 \hY_4 = s \hY_4 \,, \qquad s=\pm 1 \, .
\ee
These conditions
imply that $\hY_4$ is harmonic. 
Assume furthermore that the 4-form is homogeneous with degree $n$ on the cone.
That is
\be
\Lag_{r \pa_r} \hat Y_4 = n \hat Y_4 \,.
\ee
Here $\Lag$ denotes the Lie derivative. 
Then
we can decompose the form as
\be\label{eq:Ydecompose}
\hY_4 = r^n \left(\frac{dr}{r} \wedge Y_3 + Y_4 \right) \,,
\ee
with $Y_3$ and $Y_4$ being forms on the Sasaki-Einstein manifold $M_7$, and
(\ref{eq:conditions1}) implies
\begin{equation} \label{dY3eq}
 \star_7 d Y_3 = s n Y_3 \,, \qquad d \star_7 Y_3 = 0 \, .
\end{equation}
Now consider the 3-form fluctuation
\begin{equation} \label{deltaC}
 \delta C = \phi \, Y_3 + \text{c.c.}\, ,
\end{equation}
where $\phi$ only depends on the four dimensional spacetime coordinates. The field $\phi$ will be a pseudoscalar because the 3-form field changes sign under space or time reflections \cite{Duff:1986hr}. From (\ref{dY3eq}) one shows that in a neutral background with $A_\mu=0$ \cite{Duff:1986hr}:
\begin{equation} \label{ddphi}
 \nabla^\mu \nabla_\mu \phi = m^2 \phi \, , \qquad m^2 = \frac{n (n+6s)}{4 L^2} \, .
\end{equation}
Moreover this mode does not source any other KK modes at linear order \cite{Duff:1986hr}. 

The question now is what happens when $A_\mu$ is nonzero. We claim the following:
\vskip2mm
\emph{If $\hY_4$ is a primitive and closed (4,0) or (3,1)-form on the Calabi-Yau fourfold, and $\{g_{\mu\nu},A_\mu\}$ solve the 4d Einstein-Maxwell equations, then the covariantization of the mode (\ref{deltaC}) linearly decouples from all other Kaluza-Klein modes and satisfies the covariantized equation of motion (\ref{ddphi}), with $s=+1$ for $(4,0)$-forms and $s=-1$ for $(3,1)$-forms.}
\vskip2mm
Before sketching the proof, let us clarify the claim. Recall that a primitive middle dimensional form
on a K\"ahler manifold is one that satisfies 
\be\label{eq:primitivity}  
\hat \w \wedge \hY_4 = 0 \quad \mbox{ or equivalently } \quad \hat \w \cdot \hY_4 = 0 \,,
\ee
where $\hat \w$ is the K\"ahler form on the Calabi-Yau cone.
Covariantization means replacing, in the coordinates of
(\ref{eq:M7}), $d\psi \to d\psi + A$ in (\ref{deltaC}), and $\nabla_\mu \to \nabla_\mu - i q A_\mu$ in (\ref{ddphi}), where we assumed the mode to have a definite charge $q$ under the canonical $U(1)$ symmetry of the cone:
\be
\Lag_{\pa_\psi} \hY_4 = i q \hY_4 \,.
\ee
This charge will be directly inherited by $Y_3$ and $Y_4$. 
Thus, explicitly, we take
\be\label{eq:mode}
\delta C = \phi \, Y_3^A  + \text{c.c.} \,,
\ee
where $Y_3^A$ is obtained from $Y_3$ by replacing $d\psi$ by $d\psi + A$. In components
\be
\delta C_{mnp} = \phi \, Y_{3 \, mnp} \,, \qquad \delta C_{\mu m n} = \phi \, A_{\mu} Y_{3 \, mn \psi} \,,
\ee
where $m,n,p$ are indices on $M_7$ and $\mu$ on $M_4$.

We will discuss the existence of modes $\hY_4$ satisfying all of the above conditions in the next subsection.
For the moment we assume existence. To prove our claim, first note that the K\"ahler form on the cone may be decomposed as (see e.g. \cite{Martelli:2006yb})
\be\label{eq:w8}
\hat\w = r^2 \left( \frac{dr}{r} \wedge \eta + \w\right) \,,
\ee
where $d \eta = 2 \w$, and $\w$ is as before the K\"ahler form of $M_6$. In terms of the metric we wrote in (\ref{eq:M7}) above, $\eta = a d\psi + \sigma$.
The primitivity condition (\ref{eq:primitivity}) is easily seen to imply
\be\label{eq:Y3}
 \w \cdot Y_3 = 0 \,, \qquad \w \wedge Y_3 + s \eta \wedge \star_7 Y_3 = 0 \,.
\ee
Here we also used (\ref{eq:conditions1}). By plugging the mode (\ref{eq:mode}) into the eleven dimensional equations of motion
and using (\ref{eq:Y3}), we obtain\footnote{We will not reproduce the straightforward but tedious computations here. We verified our results using the abstract tensor calculus package {\tt xAct} \cite{xAct}.} the following three results:
\begin{itemize}
\item {\bf Decoupling from metric fluctuations}: The 3-form mode (\ref{eq:mode}) does not
source any linearised metric fluctuations provided that
\be\label{eq:metricdecouple}
F \wedge F \left(\w_n{}^q Y_{3 \, m q \psi} + \w_m{}^q Y_{3 \, n q \psi} \right) = 0 \,.
\ee

\item {\bf Decoupling from other 3-from modes}: The 3-form mode (\ref{eq:mode}) does not
source any other linearised 3-form fluctuations provided that
\be\label{eq:formdecouple}
(s+1)\, \w \wedge Y_3 = 0 \,.
\ee

\item {\bf Equation of motion for the pseudoscalar}: If decoupling occurs, then the four dimensional
pseudoscalar field satisfies
\be\label{eq:pseudo}
  \left(\nabla^\mu - i q A^\mu
 \right) \left(\nabla_\mu - i q A_\mu \right) \phi = m^2 \phi \, , \qquad m^2 = \frac{n (n+6s)}{4 L^2} \, .
\ee
Solving this equation is sufficient to solve the full 11d linearised supergravity equations.
\end{itemize}

We now proceed to characterise forms for which (\ref{eq:metricdecouple}) and (\ref{eq:formdecouple})
hold. In our electrically charged AdS-Reissner-Nordstrom background, $F \wedge F$ vanishes.
Therefore the decoupling of metric fluctuations will be automatic. However, one might certainly
wish to consider dyonic black holes also (for instance to study phenomena such as the Hall
or Nernst effects \cite{Hartnoll:2007ai, Hartnoll:2007ih, Hartnoll:2007ip}) for which this
term does not vanish. Therefore in order to solve (\ref{eq:metricdecouple}) we will require that
$\w_n{}^q Y_{3 \, m q \psi} + \w_m{}^q Y_{3 \, n q \psi}  = 0$. There are (at least) four interesting
cases in which this is true. These are if $\hY_4$ is a $(4,0)$, $(0,4)$, $(3,1)$ or $(1,3)$ form
on the eight dimensional Calabi-Yau cone.  Let us consider these cases one at a time.

If $\hY_4$ is a $(4,0)$ form, then $Y_{3 \, m q \psi}$ is zero. This follows from the fact
that $dr \wedge \eta$ is a $(1,1)$ form on the Calabi-Yau cone,
see for instance (\ref{eq:w8}). If $Y_3$ had a $d\psi$ component (i.e. an $\eta$ component),
then $\hY_4$ in (\ref{eq:Ydecompose}) would necessarily
have an antiholomorphic component and could not be $(4,0)$. Hence $Y_{3 \, m q \psi}$
is zero.

If  $\hY_4$ is a $(3,1)$ form, then $Y_{3 \, m q \psi} dx^m \wedge dx^q$ is a $(2,0)$ form.
This again follows from the decomposition of $\hY_4$ in (\ref{eq:Ydecompose})
and the fact that $dr \wedge \eta$ is a $(1,1)$ form.
Given that both $m$ and $q$ are holomorphic indices it follows that
$\w_n{}^q Y_{3 \, m q \psi} + \w_m{}^q Y_{3 \, n q \psi}  = i(Y_{3 \, m n \psi} + Y_{3 \, n m \psi})=0$.
The first of these equalities follows from the fact that $ \w_m{}^q$ is proportional to the complex structure
while the second equality follows from antisymmetry of $Y_3$.

These arguments clearly go through identically when $\hY_4$ is $(0,4)$ or $(1,3)$.
They do not work however when $\hY_4$ is a $(2,2)$ form.
We can recall at this point that $(4,0)$ forms are always primitive (from (\ref{eq:primitivity})) and self-dual whereas primitive $(3,1)$ forms are anti-self-dual, in the canonical
orientation with which we are working.

In order for (\ref{eq:formdecouple}) to vanish and other 3-form modes to decouple,
we need that either $s=-1$ or that $\w \wedge Y_3 = 0$. The first of these will hold if and
only if $\hY_4$ is anti-self-dual whereas the second holds if $\hY_4$ is a $(4,0)$ form.
This last statement follows from noting that the structure of the eight dimensional K\"ahler form (\ref{eq:w8}) implies that $\frac{dr}{r} + i \eta$ is a holomorphic 1-form on the Calabi-Yau cone. Therefore
in order for $\hY_4$ to be $(4,0)$ the decomposition
(\ref{eq:Ydecompose}) must take the form $\hY_4 = r^n (\frac{dr}{r} \pm i \eta) \wedge Y_3$,
with $Y_3$ a $(3,0)$ on the six dimensional K\"ahler-Einstein base of the Sasaki-Einstein
manifold. However, if $Y_3$ is a $(3,0)$ form, then $\w \wedge Y_3$ is zero.

This proves our claim. Summarising: The mode (\ref{eq:mode}) decouples
from all other perturbations if the closed 4-form $\hY_4$ is a $(4,0)$ or primitive
$(3,1)$-form on the Calabi-Yau cone. It is described by a minimally coupled
pseudoscalar in four dimensions with charge $q$ and mass squared
\begin{eqnarray} \label{eq:mass4} \label{eq:mass3}
L^2 m^2_{(4,0)} = \left( \frac{n}{2} + 3 \right) \frac{n}{2}  \, , \\
L^2 m^2_{(3,1)} = \frac{n}{2} \left( \frac{n}{2} - 3 \right)  \, .
\end{eqnarray}
The same expressions hold for $(0,4)$ and $(1,3)$ forms, respectively. Using the relation
$(Lm)^2 = \Delta(\Delta-3)$, we can read off the possible conformal dimensions of the dual operators.

\subsection{Existence}
\label{sec:modes}

We will now establish the existence of modes in the classes described above, and confirm that in many examples they lead to instabilities and superconductivity at low temperatures.

All Calabi-Yau cones admit a canonical holomorphic $(4,0)$ form. This form is thus closed
and self-dual. If we introduce holomorphic vielbeins $\theta^a$, $a$ runs from 1 to 4, such
that the metric is written $ds^2_{M_8} = \theta^a \bar \theta^a$, then the form is given by
\be
\hY_4 = \hat \Omega_4 \equiv \theta^1 \wedge \theta^2 \wedge \theta^3 \wedge \theta^4 \,.
\ee
It is immediate that this form has scaling dimension $n=4$ under the homothetic vector $r \pa_r$,
as the metric has scaling dimension 2 and hence the $\theta^a$ have scaling dimension 1.
Furthermore, we can easily obtain the charge $q=4 a$ by noting that for this mode
\be
\Lag_{\pa_\psi} \hY_4 = \pa_\psi \cdot d \hY_4 + d (\pa_\psi \cdot \hY_4)
 = d (\pa_\psi \cdot \hY_4) = a i \, d(r^4 Y_3) = 4 a i\, \hY_4 \,.
\ee
In the third and fourth equalities we used the fact noted previously that holomorphic 4-forms must take
the form $\hY_4 = r^n \left(\frac{dr}{r} + i \eta \right) \wedge Y_3$. In the last equality we also
used the first expression in (\ref{eq:Y3}). It is clear that this argument will apply to any closed (4,0)-form with scaling dimension $n$, giving charge $q=n a$. Such forms are readily obtained by multiplying $\hat \Omega_4$ by a homogeneous holomorphic function of degree $n-4$.

It follows from (\ref{eq:gtoa}) that all of the Sasaki-Einstein vacua have a decoupled pseudoscalar mode with charge $\gamma q = 2$ and, from (\ref{eq:mass4}), mass squared $m^2 L^2 = 10$. This corresponds to an operator of dimension $\Delta = 5$. Comparing with figure 1 or equation (\ref{eq:criterion}) we see that this mode
never leads to an instability.

The recent results of \cite{Gauntlett:2009zw} imply that this mode is part of a long vector\footnote{And hence not part of a short hypermultiplet as was claimed in \cite{Ceresole:1984hr}.} ${\rm OSp}(2|4)$ supermultiplet (the $E_0=4$, $y=0$ case in table 1 of \cite{Ceresole:1984hr}) which consistently decouples from all other Kaluza-Klein modes even at the nonlinear level. There are no other charged scalars in this multiplet.

Before moving on to consider a general class of $(3,1)$-forms, we can consider the special case
of $M_7 = S^7$ for which the Calabi-Yau cone is simply $M_8 = \C^4$. A $(3,1)$-form on $\C^4$ is given by,
for instance,
\be
\hY_4 = d\bar z_1 \wedge d z_2 \wedge d z_3 \wedge d z_4 \,.
\ee
This is a closed, primitive, anti-self-dual $(3,1)$-form with $n=4$ and $\gamma q=1$, recalling that $a=1$ for the seven sphere. From (\ref{eq:mass3}) the four dimensional mass will be $m^2 L^2 = -2$, corresponding to $\Delta=2$ or $\Delta=1$. This is precisely the value of the mass
studied in detail in \cite{Hartnoll:2008vx, Hartnoll:2008kx}.
The two different dimensions of the dual operator correspond to theories that are related via a renormalisation group flow generated by a double
trace deformation \cite{Witten:2001ua} .
From figure 1 or equation (\ref{eq:criterion}) we see that this mode does condense at low
temperatures. Therefore, {\it the IR conformal fixed point of ${\mathcal{N}}=8$ $SU(N)$ Yang-Mills theory at large $N$ spontaneously breaks $U(1)_R$ and becomes a superconductor at low temperatures and nonzero chemical potential.} Taking $\Delta=2$, we numerically find that the critical temperature is $T_c \approx 0.007 \, \mu$. For $\Delta=1$, we get $T_c \approx 0.35 \, \mu$. We recall this is a lower bound.\footnote{There is another known instability for the case of $M_7=S^7$, the Gubser-Mitra instability \cite{Gubser:2000ec, Gubser:2000mm}. That instability corresponds to the charge becoming redistributed among the more than one $U(1)$ symmetry in the theory, and does not induce superconductivity, as all the operators involved are neutral. In our units $T_\text{G-M}= \mu/\pi \approx 0.32 \mu$. Thus in the
$\Delta=1$ case the superconducting instability kicks in before the Gubser-Mitra instability.\label{gm}}

We now turn to our main source of examples, namely $(3,1)$-forms associated to complex structure moduli of the Calabi-Yau fourfold cone. Consider a metric deformation $\delta g_{ab}$, with $a$ and $b$ both holomorphic indices, preserving Ricci flatness. Then this is a Lichnerowicz zero mode and
\begin{equation} \label{zeromode}
 \hat Y_4 \equiv \delta g_{\bar a \bar e} \, {\hat{\Omega}_4^{\bar e}}{}_{bcd} \, d\bar{z}^{\bar a} \wedge dz^b \wedge dz^c \wedge dz^d \,,
\end{equation}
is a harmonic (3,1) form \cite{Candelas:1990pi}. In this equation a bar denotes an antiholomorphic index. It is easy to see that this form is furthermore primitive. We thus get an example of an anti-self-dual closed (3,1)-form, as considered above. Calabi-Yau metric deformations which preserve the cone structure (\ref{eq:M7})-(\ref{M8cone}), and are therefore moduli of the Sasaki-Einstein manifold, have the same scaling dimension as the metric and are neutral under the $U(1)$ isometry (otherwise they would not preserve the isometry and the metric would no longer be Sasaki-Einstein). Thus the associated $\hat{Y}_4$ has the same scaling dimension $n=4$ as $\hat{\Omega}_4$, and the same charge $\gamma q = 2$. The mass formula (\ref{eq:mass3}) now implies $\Delta_+=2$, saturating the BPS bound.\footnote{This mode is thus the lowest component of an OSp(2,4) hypermultiplet. Its scalar superpartner is the metric modulus fluctuation, which has $\gamma q = 0$ and $\Delta_+ = 3$, as expected for a marginal deformation.} Such modes always condense at low temperature, with (see
figure 2 above)
\begin{equation} \label{Tcmarg}
 T_c \approx 0.0416 \, \frac{\mu}{\gamma} \, .
\end{equation}
Therefore: {\it The IR fixed point of ${\mathcal{N}}=2$ $SU(N)$ Yang-Mills theories at large $N$
with Sasaki-Einstein duals with at least one metric modulus become superconducting at temperatures below (\ref{Tcmarg})}. As previously, this is a lower bound on $T_c$, there may be other unstable modes
with higher critical temperatures.

Not all Sasaki-Einstein metrics have deformation moduli. For example the round sphere has none. However, many of the Brieskorn-Pham links introduced in section \ref{sec:BPlinks} have plenty of moduli, obtained as polynomial deformations of the same weight $d$ as the original polynomial (\ref{eq:weighted}). The number of such moduli equals the number of monomials of weight $d$ minus the number of coordinate transformations respecting the weights \cite{BGbook}, that is
\begin{equation}
 N_{\rm mod} = N_{\rm mon}(d) - \sum_i N_{\rm mon}(w_i) \, ,
\end{equation}
where $N_{\rm mon}(w)$ stands for the number of monomials of weight $w$. For the example (\ref{example}), $N_{\rm mod}=1$: There is precisely one deformation which cannot be reabsorbed in a weight preserving coordinate transformation, namely $\delta F(z) = \epsilon z_3^3 z_5^4$. We shall look more systematically at the existence of moduli in the following section.

One could also consider deformations $\delta F = \epsilon F'$ of the defining equation (\ref{eq:weighted}) with weight $d' \neq d$. Such deformations do not preserve
the cone structure, and so they are not moduli of the Sasaki-Einstein space. However if the fluctuation preserves the Ricci flatness of the cone metric to linear order, then (\ref{zeromode}) still gives a harmonic (3,1) form, and the corresponding pseudoscalar mode still satisfies all the required properties to be minimally coupled. To determine the charge of such a metric fluctuation it is useful to formally associate a charge $q_\epsilon=w_\epsilon=d-d'$ to $\epsilon$. This way the polynomials $F$ and $\delta F$ would have the same charge $d$. The charge of the metric mode $\delta g_{ab} = \partial_\epsilon g_{ab} |_{\epsilon=0}$ is thus seen to be $-q_\epsilon = d'-d$. The associated form mode (\ref{zeromode}) thus has charge $q$ and radial scaling dimension $n$ given by
\begin{equation}
 \frac{n}{2} = \gamma q = 2 + \gamma (d'-d) \, .
\end{equation}
As an example, consider the deformation $\delta F = \epsilon z_1 z_2^2$ of (\ref{example}). This has $d'=756$, and so, using $\gamma=\frac{2}{113}$, we get $\frac{n}{2} = \gamma q = \frac{58}{113} \approx 0.5132$. If this truly corresponded to a Calabi-Yau preserving deformation, it would give rise to a minimally coupled BPS pseudoscalar with this value of $\Delta = \gamma q$. This leads to $T_c = 1.47318 \frac{\mu}{\gamma}$, substantially higher than the cone-preserving modes (\ref{Tcmarg}). Determining in general when such modes are indeed Calabi-Yau preserving appears to be an interesting open mathematical problem \cite{yaupriv}. We shall not address this problem here, but note that it could lead to higher values of $T_c$ than the ones we will discuss.

\subsection{Comment on the dual field theories and operators}

The gravity backgrounds that we have been describing are dual to ${\mathcal{N}}=2$ superconformal field theories. The supersymmetry and conformality follow directly from the global (super)symmetries of the gravitational solutions. In special cases there may be an enhancement of supersymmetry. For instance, when $M_7 = S^7$ the theory has ${\mathcal{N}}=8$ supersymmetry and if $M_7$ is tri-Sasakian then the theory will have ${\mathcal{N}}=3$ supersymmetry.

More specifically, the dual field theory is that describing the worldvolume dynamics of $N$ M2 branes placed at the tip of a Calabi-Yau fourfold cone over the Sasaki-Einstein manifold $M_7$ \cite{Klebanov:1998hh}. Until recently, this relationship was not useful for obtaining an explicit description of the field theory degrees of freedom. On general grounds one might expect the M2 brane theories to arise as IR fixed points of multiple D2 brane gauge theories in a background obtained by dimensionally reducing the M theory geometry along the $U(1)$ isometry of the Sasaki-Einstein metric \cite{Itzhaki:1998dd}.
This reduction will break all the manifest supersymmetry of the background for generic ($\CN=2$) Sasaki-Einstein manifolds. This occurs because the Killing spinor is charged under the $U(1)$ isometry, as we recalled below (\ref{eq:gammatop}).

A different brane construction for the case $M_7 = S^7/\Z_k$ was presented in \cite{Aharony:2008ug} (ABJM), following the renewed interest in multiple M2 brane theories initiated by \cite{Bagger:2006sk, Gustavsson:2007vu, Bagger:2007jr}. The construction involves
2 NS5 branes, $N$ D3 branes and $k$ D5 branes. Upon T dualising and lifting to M theory one obtains $N$ multiple M2 branes probing a geometry that has local $\C^4/\Z_k$ singularities. The brane construction allowed \cite{Aharony:2008ug}  to identify the field theory as a specific superconformal $U(N) \times U(N)$ Chern-Simons theory at levels $k$ and $-k$.

The ABJM brane construction was generalised to a family of ${\mathcal{N}}=3$ field theories
in \cite{Jafferis:2008qz}. These are dual to backgrounds in which $M_7$ is a tri-Sasakian manifold.
The field theory dual for general ${\mathcal{N}}=2$ theories is not yet available, it appears that the most
tractable subset of ${\mathcal{N}}=2$ theories are those in which the Calabi-Yau cone $M_8$ is toric
(the Brieskorn-Pham cones we considered above are generally not toric).
Combining the extensive intuition gained from toric ${\mathcal{N}}=1$ superconformal field theories in 3+1 dimensions and the ABJM construction, it has been proposed that the worldvolume theory of M2 branes probing toric Calabi-Yau cones is given by a quiver Chern-Simons theory
\cite{Martelli:2008rt, Martelli:2008si, Hanany:2008cd,Ueda:2008hx,Imamura:2008qs,Hanany:2008fj}. These have large gauge symmetries
with associated gauge fields $A_i$ and complex scalar fields $\phi_a$ that are charged under
the gauge symmetries. The supermultiplets are then completed with additional scalar and spinor
fields. The action takes the form
\bea
S & = & \sum_i \frac{k_i}{4\pi} \int d^3x \Tr \left( A_i \wedge d A_i + \frac{2}{3} A_i \wedge A_i \wedge A_i + \text{superpartners} \right) \nonumber \\
 & + & \sum_a \int d^3x \left( \left| D \phi_a \right|^2  - \left| \frac{\pa W}{\pa \phi_a} \right|^2 + \text{superpartners} \right) \,.
\eea
We are being somewhat schematic. The superpotential $W$ is a holomorphic function of the $\phi_a$.
A thorough discussion of ${\mathcal{N}}=2$ Chern-Simons
theories may be found in \cite{Gaiotto:2007qi}. The point we would like to emphasise is
that concrete field theory duals have been proposed for certain Sasaki-Einstein manifolds.
One can therefore hope to identify the precise operator $\ocal$ which condenses at the
superconducting instability.

The first mode we discussed in section \ref{sec:modes} was obtained from the canonical
holomorphic $(4,0)$-form on the Calabi-Yau cone. Although this mode did not lead to an instability,
it is instructive to consider its dual field theory operator. The mode must be dual to a
canonical operator which is present in all ${\mathcal{N}}=2$ theories. The most
natural candidate is the superpotential itself: $\ocal = W$.\footnote{An analogous identification is implicitly made in the AdS$_5$/CFT$_4$ case with ${\mathcal{N}}=1$ supersymmetry in 3+1 dimensions in, for instance, \cite{Martelli:2006yb}.} As well as being holomorphic and
canonical, this mode had charge $\gamma q = 2$ which is also
the R-charge of the superpotential. However, the mode has
dimension $\Delta=5$, whereas the superpotential has classical dimension $\Delta = 2$,
as a consequence of being chiral. This identification would therefore imply that the
dimension of the superpotential is renormalised in these 2+1 theories. This is consistent with the fact, mentioned in section \ref{sec:modes}, that this mode is part of a long multiplet \cite{Gauntlett:2009zw}, so its dimension is not protected.

The second set of modes we discussed were $(3,1)$-forms corresponding
to complex moduli deformations of the Calabi-Yau cone.
These must be canonically dual to deformations of the field theory
that preserve supersymmetry and conformality. The most natural candidate dual
operators are deformations of the superpotential, $\ocal = \delta W$.
In this case our bulk mode was BPS, with charge and dimension
$\Delta = \gamma q = 2$, equal to those of bare superpotentials.
These are relevant charged operators.
This identification would indicate that whereas the overall
superpotential is renormalised, deformations of the superpotential (if they
exist) are not.

We also noted in section \ref{sec:modes} that the $(3,1)$-form modes lie
in a hypermultiplet which contained a scalar superpartner
with $\gamma q = 0$ and $\Delta=3$. This mode will be dual
to a marginal deformation of the Lagrangian.
If our previous identification with deformations of the superpotential
is correct, these operators will be of the form
$\ocal = \int d^2 \theta \, \delta W + c.c. = \partial_{\phi^a} \partial_{\phi^b} \delta W \psi^a \psi^b
+ \cdots$, with $\psi^a$ fermionic superpartners of the $\phi^a$.

It is certainly of interest to flesh out these identifications further for
cases in which the superpotential and its deformations are known
explicitly. We will leave this for future work.

\subsection{Comment on skew-whiffing}

Given a (neutral) Freund-Rubin compactification from eleven to four dimensions,
a different solution may be constructed by skew-whiffing \cite{Duff:1986hr}.
One way to describe the skew-whiffed solution is to change the sign of the
3-form background with everything else held fixed. In terms of the ansatz
(\ref{eq:ansatz}) and (\ref{eq:ansatz2}), with $A=0$, this corresponds to
letting $L \to -L$. In terms of brane constructions, this means that instead of
$N$ M2 branes at the tip of a Calabi-Yau cone, one takes $N$ anti-M2 branes.
This operation is not as innocuous as it might seem.
With the exception of the case $M_7 = S^7$, only one of the two
solutions can be supersymmetric \cite{Duff:1986hr}. At the strict classical
level, skew-whiffed solutions obtained from supersymmetric Freund-Rubin
compactifications give examples of stable non-supersymmetric
vacua \cite{Duff:1986hr}. Stability beyond the classical level is not known.

In the skew-whiffed backgrounds ($L \to -L$) it turns out that the construction of
section \ref{sec:freund} above does not give a consistent reduction
to Einstein-Maxwell theory in general. This is because a relative
sign changes between the kinetic and Chern-Simons term in the 3-form
equations of motion. However, for a purely electric (or purely magnetic)
background, such as the AdS-Reissner-Nordstrom black holes of interest
to us, the Chern-Simons term vanishes and one does obtain a
solution.

Perturbing the skew-whiffed charged background by our mode (\ref{eq:mode}) one
finds that both the decoupling conditions (\ref{eq:metricdecouple})
and (\ref{eq:formdecouple}) and the equation of motion for the pseudoscalar
(\ref{eq:pseudo}) are changed by $s \to -s$. It follows from our previous arguments that only
the modes obtained from closed $(4,0)$ and $(0,4)$ forms on the Calabi-Yau cone
decouple in this case. Their mass squared is now given by
\be
m^2_{(4,0)} = m^2_{(0,4)} = \frac{n (n-6)}{4 L^2} \,.
\ee
We recalled above that all Calabi-Yau cones admit a closed $(4,0)$ form
with $n=4$ and charge $\gamma q=2$. Therefore, all of the skew-whiffed
backgrounds have a minimally coupled pseudoscalar with
$m^2=-2$, corresponding to $\Delta=2$ or $\Delta=1$. We noted above (see figure 1) that
these values of the charge and $\Delta$ lead to a superconducting instability
at low temperatures. Therefore {\it all theories dual to skew-whiffed
Sasaki-Einstein compactifications of M theory are superconducting at low temperatures
when placed at a finite chemical potential}.

\section{A distribution of critical temperatures}

In this section we consider Sasaki-Einstein manifolds obtained
as Brieskorn-Pham links, as discussed in section \ref{sec:BPlinks}, and which
have unstable 3-form modes of the type considered in section \ref{sec:modes}.
For these theories, all the quantities in the four dimensional action (\ref{eq:action})
can be explicitly computed. This allows us to obtain a distribution of
critical temperatures.

More specifically, we will focus on the 3-form modes associated to metric moduli. Their critical temperature $T_c$ is given by (\ref{Tcmarg}).\footnote{Although we will only consider the distribution of the critical temperatures for this particular mode, we should keep in mind that there may be other modes that become unstable at higher temperatures.} Notice that $T_c/\mu$ is proportional to $\gamma^{-1}$, with constant of proportionality independent of the theory.  Therefore, in order
to obtain a distribution of critical temperatures $T_c$ at fixed $\mu$, it is
sufficient to obtain a distribution of values of $\gamma^{-1} = 2 a  \in \half \Z^+$.
We shall now note various features of this distribution for Brieskorn-Pham
cones, putting aside momentarily the question of whether or not the manifolds
have metric moduli.

The lowest value of $a$ is clearly $a=1/4$. From (\ref{Tcmarg}), this corresponds
to $T_c \approx 0.0208 \, \mu$. To gain some intuition for this result,
it is useful to express this relation in physical units.
The only quantity that we need to reintroduce is the Boltzmann constant
$k_B = 8.617 \times 10^{-5} \text{eV K}^{-1}$, which we have thus far set to unity.
Furthermore we recall that one Volt is $\text{V} = \text{eV e}^{-1}$ and that we have
set the fundamental charge $e=1$.\footnote{If these theories were to be realised in a lab,
the identification $e=1$ would only be correct if the unit of charge in the (emergent)
CFT coincided with the (standard model) electron charge.}
The lowest critical temperature we find is therefore
\be\label{eq:physicalTc}
\left. \frac{T_c}{\text{[K]}} \right|_{\text{min.}} \approx \; \; 0.241 \, \frac{\mu}{\text{[mV]}} \,.
\ee
Thus, for instance, if we put the membrane CFT at a chemical potential of one milliVolt, the critical temperature would be $0.24$ degrees Kelvin. If (\ref{eq:physicalTc}) is taken literally, then by
increasing the chemical potential we can make $T_c$ arbitrarily high.
Of course, in actual theories arising
at quantum critical points in a real-life crystal, other factors
such as impurities and interactions with background ions
would influence the onset of superconductivity.

Less obviously, there is also an upper bound on $a$ and hence
an upper bound on the critical temperatures within this class of
Sasaki-Einstein duals:
\vspace{0.2cm}

\noindent {\bf Lemma 1:} For the Brieskorn-Pham links  constructed in \cite{Boyer:2003pe} and reviewed in section
\ref{sec:BPlinks}, the metric coefficient
$a$ has an upper bound. Thus the critical temperature at fixed chemical
potential (\ref{Tcmarg}) is bounded above in these models.
\vspace{0.2cm}

\noindent This proof of this result is in Appendix
\ref{eq:proofs}. The largest value of $a$ that we found by
scanning numerically (over $m_i < 100$)
is $a=2039/4$. However, this manifold does not have moduli.
The largest value of $a$ we found for a manifold with moduli
is $a=683/4$. The defining polynomial (\ref{eq:fermat}) for this case is
$F = z_1^2+z_2^3+z_3^7 + z_4^{37} + z_5^{99}$.
There is a single modulus $\delta F = z_2^2 z_5^{33}$.
This value of $a$ leads to $T_c \approx 14.2 \mu$.
Introducing physical units as above leads to
\be
\left. \frac{T_c}{\text{[K]}} \right|_{\text{max.}} \approx \; \; 165 \, \frac{\mu}{\text{[mV]}} \,.
\ee
Thus $T_c$ is $165$ Kelvin if the system is at a chemical potential of one milliVolt.
This is likely not the maximum $T_c$ attainable, rather it is
the largest value we found by scanning numerically.

A second interesting result is that while there
are infinitely many Brieskorn-Pham links that lead to
Sasaki-Einstein manifolds, only a handful
of values of $a$ occur infinitely many times.
\vspace{0.2cm}

\noindent {\bf Lemma 2:} There are precisely 19 values of $a$
which occur infinitely many times in the Brieskorn-Pham links. These
are
\be
a = \frac{n}{4} \,,
\ee
where $n = \{1,2,3,4,5,6,7,8,9,10,12,14,15,18,20,21,24,30,42\}$.
\vspace{0.2cm}

\noindent The proof of this result is again in Appendix
\ref{eq:proofs}. This series suggests the resolution of a
puzzle raised in \cite{Adams:1979wf}.

It is straightforward to scan numerically through different values
of the exponents $\{m_i\}$ in the defining polynomials for the
Brieskorn-Pham cones, and to check whether they satisfy the condition
(\ref{eq:condition}) for being Sasaki-Einstein.
We then need to check whether the Sasaki-Einstein
space has metric moduli.
Each time we find a solution with metric moduli,
we can compute $a$ and hence $T_c$, via (\ref{Tcmarg}).  In Figure 3 we show the
solutions obtained for a scan over all $m_i < 100$.
This scan led to
7278 distinct Sasaki-Einstein manifolds, 6190 of which had metric moduli.
As noted below (\ref{eq:gammatop}) above, we can also consider quotients of
these manifolds by $\Z_k$, with $k$ a divisor of $4a$. After considering
quotients of the manifolds with moduli, we obtain 11,821 solutions.
The critical temperatures of these manifolds are shown in Figure 3.
Of the 7278 manifolds found, only around
350 belong to the infinite families of theorem 2. Removing them does not
change the distribution noticeably. It seems therefore that figure
3 accurately captures the distribution of critical temperatures in the
finitely many theories which do not belong to infinite families.

\begin{figure}[h]
\label{fig:distribution}
\begin{center}
\includegraphics[height=9cm]{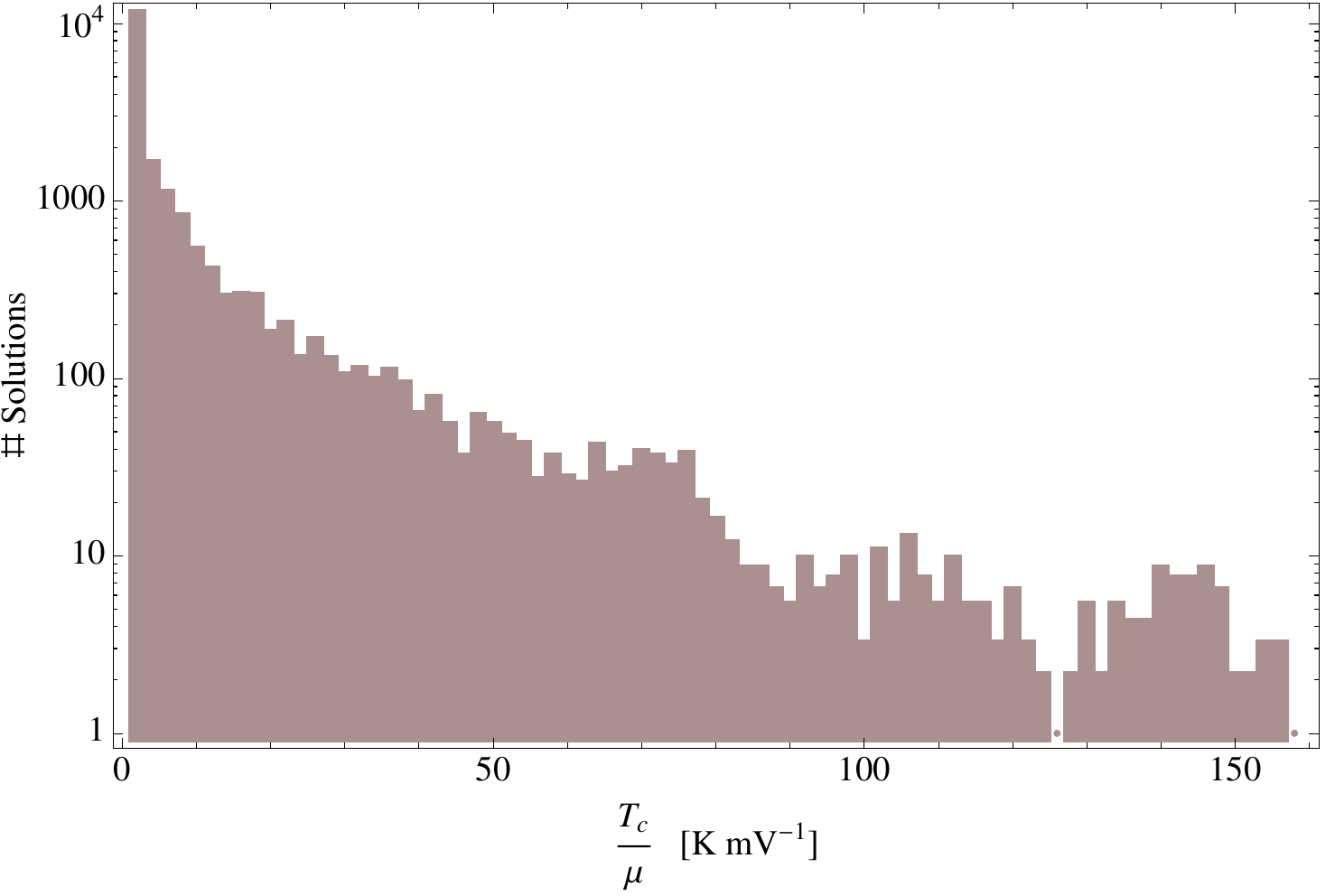}
\end{center}
\caption{A logarithmic distribution of critical temperatures over the chemical potential,
in units of degrees Kelvin per milliVolt. The distribution is obtained from
a scan over Brieskorn-Pham cones admitting Sasaki-Einstein metrics with moduli,
along with allowed $\Z_k$ quotients.
The solutions have been binned into ranges of width $2$ K/mV.}
\end{figure}

In figure 3 we see that the critical temperatures cluster around
the lowest value $T_c/\mu \approx 0.241\, \text{K/mV}$. The clustering appears
to roughly follow a power law.

We close this discussion by noting that the instability we found
for the maximally supersymmetric (${\mathcal{N}}=8$) theory in section
\ref{sec:modes}, which is not due to a modulus mode and not included
in figure 3, gives the following critical temperatures
in physics units:
\be
\left. \frac{T_c}{\text{[K]}} \right|_{{\mathcal{N}}=8} \approx \; \; 0.081 \, \frac{\mu}{\text{[mV]}} \quad
\text{or} \quad 4.1 \, \frac{\mu}{\text{[mV]}} \,,
\ee
corresponding to $\Delta=2$ and $\Delta=1$ for the operator that condenses, respectively.
We noted in footnote \ref{gm} above that in the $\Delta=1$ case, this instability
occurs at a higher temperature than the Gubser-Mitra instability of
the ${\mathcal{N}}=8$ theory at a finite chemical potential.

\section{Discussion}

In this paper we have given the first explicit string theory
realisations of the onset of an s wave superconducting phase
in strongly coupled field theories at finite chemical potential
as considered in \cite{Gubser:2008px, Hartnoll:2008vx, Hartnoll:2008kx}.
The main technical result that made this possible
was the identification of charged modes in Sasaki-Einstein
compactifications which decoupled from other modes
at a linearised level, even in the presence of a background
Maxwell field. Our results sit at the intersection of three directions of current
string theory research: the string landscape, AdS/CFT duality
for M2 brane theories and applications of AdS/CFT to condensed
matter physics. This leads to future research questions with
differing flavours.

In order to obtain a more complete picture of the
superconducting physics of these ${\mathcal{N}} \geq 2$ theories
there are two important questions we have not addressed.
Still at the linearised level, one should perform a complete
stability analysis with all of the coupled scalar, vector and
tensor modes. This way one can identify the most unstable mode,
obtain the precise critical temperature and determine whether
or not {\it all} Sasaki-Einstein compactifications become
superconducting. If the most unstable mode is a
charged vector or tensor, one might obtain
p wave (cf. \cite{Gubser:2008zu, Gubser:2008wv, Roberts:2008ns})
or d wave superconductors, respectively.
Beyond the linearised level, one would ultimately like
to find the endpoint of the instability well below $T_c$. These will be hairy
black hole solutions of M theory. Given the full solution there
will be many properties to investigate, starting with the
possible existence of a mass gap.

The recent progress in constructing field theory duals
to AdS$_4$ backgrounds opens various interesting
future directions. One would like to identify precisely
the operators which condense
and ultimately gain some dynamical understanding
of what is driving the instability.
Also, if the field theory admits a weak coupling
limit, one can ask whether the superconducting phase
continues to weak coupling. In fact, it is rather
natural that a weakly coupled theory with massless charged
bosonic degrees of freedom become superconducting
when placed at a finite chemical potential. This is because
the chemical potential acts as a negative mass squared.
It would also be interesting, therefore, if there are theories that
are superconducting at weak coupling but not strong coupling.

In terms of field theory duals, one is not restricted to
AdS$_4$/CFT$_3$. It seems likely that Sasaki-Einstein
compactifications to AdS$_5$ will have similar
instabilities. If so, this will lead to superconducting
phases in very well studied field theories with AdS$_5$ duals.
It was checked in \cite{Horowitz:2008bn}
that the basic mechanism of holographic superconductivity
generalises to AdS$_5$.

Regarding the string landscape; we have considered here
only the simplest (Freund-Rubin) flux compactifications of string/M theory. As we
noted in section \ref{sec:weak}, the logic behind the weak gravity
bound, if correct, suggests that theories dual to generic AdS$_4$ flux
compactifications should have a superconducting phase
when considered at a finite chemical potential. A natural question is to
scan the wider string theory landscape in search of superconductors.
As in this work, the main technical difficulty will be to identify a sector for which the stability analysis becomes tractable.

In the introduction we highlighted a parallel between the string landscape and
the atomic landscape of condensed matter physics. It would be
fascinating if this connection could be made literal by actually engineering
a (large $N$) supersymmetric gauge theory in a lab. Emergent gauge fields
are known to occur in certain lattice systems, see e.g. \cite{wen}.
One conceptually interesting consequence of such a connection would be that a
standard model lattice vacuum would provide a non-perturbative
definition of string theory (with specific AdS asymptotic boundary conditions),
thus inverting the traditional roles of string theory and the standard model.

\section*{Acknowledgements}

We are happy to acknowledge helpful discussions with many people
while this work has been in production. In particular
Dionysis Anninos, Nima Arkani-Hamed, Jonathan Bagger, David Berenstein, Charles Boyer, Miranda Cheng, Michael Duff, Mboyo Esole,
Amihay Hanany, Matt Headrick, Igor Klebanov, Don Marolf, Nai-Phuan Ong, Matthew Roberts, Subir Sachdev,
James Sparks, Erik Verlinde, S.T.~Yau, and especially Alessandro Tomasiello and Xi Yin.
We would also like to acknowledge the hospitality of Monsoon workshop at the TIFR in Mumbai
while this work was underway.
This work is supported in part by DOE grant
DE-FG02-91ER40654 and a DOE OJI award.

\appendix

\section{Proofs for the distribution of conductivities}
\label{eq:proofs}

\noindent {\bf Lemma 1:} For the Brieskorn-Pham links,
\be\label{eq:c}
a = \frac{\text{lcm}(m_i | i = 1..5)}{4} \left(\sum \frac{1}{m_i}
- 1 \right) \,,
\ee
has an upper bound.
\vspace{0.1cm}

\noindent {\bf Proof:} For $a$ to be unbounded, clearly at least
one of the $m_i$, call it $m_5$, must become arbitrarily large.

Suppose that $\sum_{i=1}^4 1/m_i < 1$. It can be shown
\cite{Boyer:2003qy} that given that the $m_i$ are positive
integers, this requires $\sum_{i=1}^4 1/m_i \leq 1805/1806$. The
first inequality in (\ref{eq:condition}) now requires that $1/1806
< 1/m_5$, and hence $m_5 < 1806$ is bounded.

Suppose instead that $\sum_{i=1}^4 1/m_i = 1 + X$, with $X \geq
0$. The second inequality in (\ref{eq:condition}) implies that $X
< 1/(3 m_5)$. We show a couple of paragraphs down that for $i \neq
5$ we must have $m_i \leq 42$. It follows that if $X>0$, then $X$
cannot be made arbitrarily small, and hence $m_5 < 1/(3 X)$ gives
a bound for $m_5$.

The remaining case to consider is $X=0$, that is, $\sum_{i=1}^4
1/m_i = 1$. Here $m_5$ is not bounded. However, the formula
(\ref{eq:c}) for $a$ in this case implies that $4 a \leq m_1 m_2 m_3
m_4$. Because $m_i \leq 42$, for $i \neq 5$, then this is bounded.

To complete the proof we need to show that $m_i \leq 42$, for $i
\neq 5$, when $\sum_{i=1}^4 1/m_i \geq 1$. Firstly, note that
$\sum_{i=1}^3 1/m_i < 1$ because otherwise the second inequality
in (\ref{eq:condition}) implies $1/m_4 + 1/m_5 < 4/(3 m_5) <
2/m_5$ which contradicts the fact that $m_5 \geq m_4$. From this
inequality it can be shown \cite{Boyer:2003qy} that $\sum_{i=1}^3
1/m_i \leq 41/42$. Combining this fact with $\sum_{i=1}^4 1/m_i
\geq 1$ implies that $m_4 \leq 42$. Swapping the labels around, this
argument gives $m_1, m_2, m_3, m_4 \leq 42$, as required.

\qed

\noindent {\bf Lemma 2:} There are precisely 19 values of $a$
which occur infinitely many times in the Brieskorn-Pham links. These
are
\be
a = \frac{n}{4} \,,
\ee
where $n = \{1,2,3,4,5,6,7,8,9,10,12,14,15,18,20,21,24,30,42\}$.
\vspace{0.1cm}

\noindent {\bf Proof:} We noted in the proof of our previous
lemma that the largest exponent $m_5$ can only become unbounded
if: $\sum_{i=1}^4 1/m_i = 1$. However, there are only 14 different
sets of $(m_1,m_2,m_3,m_4)$ for which this is possible. Namely:
$(2,3,7,42)$, $(2,3,8,24)$, $(2,3,9,18)$, $(2,3,10,15)$,
$(2,3,12,12)$, $(2,4,5,20)$, $(2,4,6,12)$, $(2,4,8,8)$,
$(2,5,5,10)$, $(2,6,6,6)$, $(3,3,4,12)$, $(3,3,6,6)$, $(3,4,4,6)$,
$(4,4,4,4)$. For sufficiently large integer $k$, any of these sets
together with $m_5 = k$ solves the conditions
(\ref{eq:condition}). It is then simple to use the formula
(\ref{eq:c}) to obtain the 19 values of $n$ that appear in the
statement of this theorem.

\qed


\begin{thebibliography}{99}

\bibitem{Maldacena:1997re}
J.~M.~Maldacena, ``The large N limit of superconformal field
theories and supergravity,'' Adv.\ Theor.\ Math.\ Phys.\ {\bf 2}
(1998) 231 [Int.\ J.\ Theor.\ Phys.\ {\bf 38} (1999) 1113]
[arXiv:hep-th/9711200].

\bibitem{Bousso:2000xa}
 R.~Bousso and J.~Polchinski,
 ``Quantization of four-form fluxes and dynamical neutralization of the
 cosmological constant,''
 JHEP {\bf 0006}, 006 (2000)
 [arXiv:hep-th/0004134].

\bibitem{Kachru:2003aw}
 S.~Kachru, R.~Kallosh, A.~Linde and S.~P.~Trivedi,
 ``De Sitter vacua in string theory,''
 Phys.\ Rev.\  D {\bf 68}, 046005 (2003)
 [arXiv:hep-th/0301240].

\bibitem{Susskind:2003kw}
 L.~Susskind,
 ``The anthropic landscape of string theory,''
 arXiv:hep-th/0302219.

\bibitem{Douglas:2003um}
 M.~R.~Douglas,
 ``The statistics of string / M theory vacua,''
 JHEP {\bf 0305}, 046 (2003)
 [arXiv:hep-th/0303194].

\bibitem{Douglas:2006es}
 M.~R.~Douglas and S.~Kachru,
 ``Flux compactification,''
 Rev.\ Mod.\ Phys.\  {\bf 79}, 733 (2007)
 [arXiv:hep-th/0610102].

\bibitem{Denef:2008wq}
 F.~Denef,
 ``Les Houches Lectures on Constructing String Vacua,''
 arXiv:0803.1194 [hep-th].

\bibitem{Greiner}
 M.~Greiner and S.~F\"olling,
 ``Condensed-matter physics: Optical lattices,''
 Nature {\bf 453}, 736 (2008).

\bibitem{sachdev}
S. Sachdev, {\it Quantum Phase Transitions}, CUP, 1999.

\bibitem{sachdev2}
S. Sachdev, ``Quantum magnetism and criticality,''
Nature Physics 4, 173 - 185 (2008)
[arXiv:0711.3015 [cond-mat.str-el]].

\bibitem{Aharony:2002up}
 O.~Aharony,
 ``The non-AdS/non-CFT correspondence, or three different paths to QCD,''
 arXiv:hep-th/0212193.

\bibitem{Son:2008ye}
 D.~T.~Son,
 ``Toward an AdS/cold atoms correspondence: a geometric realization of the
 Schroedinger symmetry,''
 Phys.\ Rev.\  D {\bf 78}, 046003 (2008)
 [arXiv:0804.3972 [hep-th]].

\bibitem{Balasubramanian:2008dm}
 K.~Balasubramanian and J.~McGreevy,
 ``Gravity duals for non-relativistic CFTs,''
 Phys.\ Rev.\ Lett.\  {\bf 101}, 061601 (2008)
 [arXiv:0804.4053 [hep-th]].

\bibitem{Kachru:2008yh}
 S.~Kachru, X.~Liu and M.~Mulligan,
 ``Gravity Duals of Lifshitz-like Fixed Points,''
 Phys.\ Rev.\  D {\bf 78}, 106005 (2008)
 [arXiv:0808.1725 [hep-th]].

\bibitem{Witten:1998qj}
 E.~Witten,
 ``Anti-de Sitter space and holography,''
 Adv.\ Theor.\ Math.\ Phys.\  {\bf 2}, 253 (1998)
 [arXiv:hep-th/9802150].

\bibitem{Gubser:1998bc}
 S.~S.~Gubser, I.~R.~Klebanov and A.~M.~Polyakov,
 ``Gauge theory correlators from non-critical string theory,''
 Phys.\ Lett.\  B {\bf 428}, 105 (1998)
 [arXiv:hep-th/9802109].

\bibitem{Gubser:2008px}
 S.~S.~Gubser,
 ``Breaking an Abelian gauge symmetry near a black hole horizon,''
 arXiv:0801.2977 [hep-th].

\bibitem{Hartnoll:2008vx}
 S.~A.~Hartnoll, C.~P.~Herzog and G.~T.~Horowitz,
 ``Building a Holographic Superconductor,''
 Phys.\ Rev.\ Lett.\  {\bf 101}, 031601 (2008)
 [arXiv:0803.3295 [hep-th]].

\bibitem{Hartnoll:2008kx}
 S.~A.~Hartnoll, C.~P.~Herzog and G.~T.~Horowitz,
 ``Holographic Superconductors,''
 arXiv:0810.1563 [hep-th].

\bibitem{Gubser:2008zu}
  S.~S.~Gubser,
  ``Colorful horizons with charge in anti-de Sitter space,''
  Phys.\ Rev.\ Lett.\  {\bf 101}, 191601 (2008)
  [arXiv:0803.3483 [hep-th]].

\bibitem{Gubser:2008wv}
  S.~S.~Gubser and S.~S.~Pufu,
  ``The gravity dual of a p-wave superconductor,''
  JHEP {\bf 0811}, 033 (2008)
  [arXiv:0805.2960 [hep-th]].

\bibitem{Roberts:2008ns}
  M.~M.~Roberts and S.~A.~Hartnoll,
  ``Pseudogap and time reversal breaking in a holographic superconductor,''
  JHEP {\bf 0808}, 035 (2008)
  [arXiv:0805.3898 [hep-th]].

\bibitem{Ammon:2008fc}
  M.~Ammon, J.~Erdmenger, M.~Kaminski and P.~Kerner,
  ``Superconductivity from gauge/gravity duality with flavor,''
  arXiv:0810.2316 [hep-th].
  
\bibitem{Basu:2008bh}
  P.~Basu, J.~He, A.~Mukherjee and H.~H.~Shieh,
  ``Superconductivity from D3/D7: Holographic Pion Superfluid,''
  arXiv:0810.3970 [hep-th].

\bibitem{Cardy:1987dg}
 J.~L.~Cardy,
 ``Anisotrpoic corrections to correlation functions in finite
 size systems,''
 Nucl.\ Phys.\ B {\bf 290}, 355 (1987).

\bibitem{Kovtun:2008kw}
 P.~Kovtun and A.~Ritz,
 ``Black holes and universality classes of critical points,''
 Phys.\ Rev.\ Lett.\  {\bf 100}, 171606 (2008)
 [arXiv:0801.2785 [hep-th]].

\bibitem{Herzog:2007ij}
 C.~P.~Herzog, P.~Kovtun, S.~Sachdev and D.~T.~Son,
 ``Quantum critical transport, duality, and M-theory,''
 Phys.\ Rev.\  D {\bf 75}, 085020 (2007)
 [arXiv:hep-th/0701036].

\bibitem{Klebanov:1999tb}
 I.~R.~Klebanov and E.~Witten,
 ``AdS/CFT correspondence and symmetry breaking,''
 Nucl.\ Phys.\  B {\bf 556}, 89 (1999)
 [arXiv:hep-th/9905104].

\bibitem{Kovtun:2008kx}
 P.~Kovtun and A.~Ritz,
 ``Universal conductivity and central charges,''
 arXiv:0806.0110 [hep-th].

\bibitem{Romans:1991nq}
  L.~J.~Romans,
  ``Supersymmetric, cold and lukewarm black holes in cosmological
  Einstein-Maxwell theory,''
  Nucl.\ Phys.\  B {\bf 383}, 395 (1992)
  [arXiv:hep-th/9203018].

\bibitem{Gibbons:2002pq}
 G.~Gibbons and S.~A.~Hartnoll,
 ``A gravitational instability in higher dimensions,''
 Phys.\ Rev.\  D {\bf 66}, 064024 (2002)
 [arXiv:hep-th/0206202].

\bibitem{Gubser:2008pf}
 S.~S.~Gubser and A.~Nellore,
 ``Low-temperature behavior of the Abelian Higgs model in anti-de Sitter
 space,''
 arXiv:0810.4554 [hep-th].

\bibitem{ArkaniHamed:2006dz}
 N.~Arkani-Hamed, L.~Motl, A.~Nicolis and C.~Vafa,
 ``The string landscape, black holes and gravity as the weakest force,''
 JHEP {\bf 0706}, 060 (2007)
 [arXiv:hep-th/0601001].

\bibitem{Duff:1986hr}
  M.~J.~Duff, B.~E.~W.~Nilsson and C.~N.~Pope,
  ``Kaluza-Klein Supergravity,''
  Phys.\ Rept.\  {\bf 130}, 1 (1986).
  
\bibitem{Gauntlett:2007ma}
  J.~P.~Gauntlett and O.~Varela,
  ``Consistent Kaluza-Klein Reductions for General Supersymmetric AdS
  Solutions,''
  Phys.\ Rev.\  D {\bf 76}, 126007 (2007)
  [arXiv:0707.2315 [hep-th]].

\bibitem{Gibbons:2002th}
 G.~W.~Gibbons, S.~A.~Hartnoll and C.~N.~Pope,
 ``Bohm and Einstein-Sasaki metrics, black holes and cosmological event
 horizons,''
 Phys.\ Rev.\ D {\bf 67}, 084024 (2003)
 [arXiv:hep-th/0208031].

\bibitem{Boyer:2004fc}
 C.~P.~Boyer and K.~Galicki,
 ``Sasakian Geometry, Hypersurface Singularities, and Einstein Metrics,''
 arXiv:math/0405256.

\bibitem{Boyer:2008vf}
  C.~P.~Boyer,
  ``Sasakian Geometry: The Recent Work of Krzysztof Galicki,''
  arXiv:0806.0373 [math.DG].

\bibitem{Martelli:2006yb}
 D.~Martelli, J.~Sparks and S.~T.~Yau,
 ``Sasaki-Einstein manifolds and volume minimisation,''
 arXiv:hep-th/0603021.

\bibitem{Bergman:2001qi}
 A.~Bergman and C.~P.~Herzog,
 ``The volume of some non-spherical horizons and the AdS/CFT
 correspondence,''
 JHEP {\bf 0201}, 030 (2002)
 [arXiv:hep-th/0108020].

\bibitem{Gauntlett:2006vf}
  J.~P.~Gauntlett, D.~Martelli, J.~Sparks and S.~T.~Yau,
  ``Obstructions to the existence of Sasaki-Einstein metrics,''
  Commun.\ Math.\ Phys.\  {\bf 273}, 803 (2007)
  [arXiv:hep-th/0607080].

\bibitem{Boyer:2003pe}
 C.~P.~Boyer, K.~Galicki and J.~Kollar,
 ``Einstein Metrics on Spheres,''
 Ann Math {\bf 162}, 557 (2005)
 [arXiv:math/0309408].

\bibitem{Boyer:2003ch}
 C.~P.~Boyer and K.~Galicki,
 ``Einstein Metrics on Rational Homology Spheres,''
 arXiv:math/0311355.

\bibitem{Boyer:2003qy}
 C.~P.~Boyer, K.~Galicki, J.~Kollar and E.~Thomas,
 ``Einstein Metrics on Exotic Spheres in Dimensions 7, 11, and 15,''
 arXiv:math/0311293.

\bibitem{xAct} The xAct package is developed by Jos\'e Mart\'in-Garc\'ia and can be downloaded from \href{http://metric.iem.csic.es/Martin-Garcia/xAct/index.html}{http://metric.iem.csic.es/Martin-Garcia/xAct/index.html}.

\bibitem{Hartnoll:2007ai}
  S.~A.~Hartnoll and P.~Kovtun,
  ``Hall conductivity from dyonic black holes,''
  Phys.\ Rev.\  D {\bf 76}, 066001 (2007)
  [arXiv:0704.1160 [hep-th]].

\bibitem{Hartnoll:2007ih}
  S.~A.~Hartnoll, P.~K.~Kovtun, M.~Muller and S.~Sachdev,
  ``Theory of the Nernst effect near quantum phase transitions in condensed
  matter, and in dyonic black holes,''
  Phys.\ Rev.\  B {\bf 76}, 144502 (2007)
  [arXiv:0706.3215 [cond-mat.str-el]].

\bibitem{Adams:1979wf} 
 D.~Adams, ``The Hitchhiker's Guide to the Galaxy'', Ballantine Books, 1979.

\bibitem{Hartnoll:2007ip}
  S.~A.~Hartnoll and C.~P.~Herzog,
  ``Ohm's Law at strong coupling: S duality and the cyclotron resonance,''
  Phys.\ Rev.\  D {\bf 76}, 106012 (2007)
  [arXiv:0706.3228 [hep-th]].

\bibitem{Gauntlett:2009zw}
  J.~P.~Gauntlett, S.~Kim, O.~Varela and D.~Waldram,
  ``Consistent supersymmetric Kaluza--Klein truncations with massive modes,''
  arXiv:0901.0676 [hep-th].

\bibitem{Ceresole:1984hr}
  A.~Ceresole, P.~Fre and H.~Nicolai,
  ``Multiplet Structure And Spectra Of N=2 Supersymmetric Compactifications,''
  Class.\ Quant.\ Grav.\  {\bf 2}, 133 (1985).

\bibitem{Witten:2001ua}
  E.~Witten,
  ``Multi-trace operators, boundary conditions, and AdS/CFT correspondence,''
  arXiv:hep-th/0112258.

\bibitem{Gubser:2000ec}
  S.~S.~Gubser and I.~Mitra,
  ``Instability of charged black holes in anti-de Sitter space,''
  arXiv:hep-th/0009126.

\bibitem{Gubser:2000mm}
  S.~S.~Gubser and I.~Mitra,
  ``The evolution of unstable black holes in anti-de Sitter space,''
  JHEP {\bf 0108}, 018 (2001)
  [arXiv:hep-th/0011127].

\bibitem{Candelas:1990pi}
  P.~Candelas and X.~de la Ossa,
  ``Moduli space of Calabi-Yau manifolds,''
  Nucl.\ Phys.\  B {\bf 355}, 455 (1991).

\bibitem{BGbook}
 C.~P.~Boyer and K.~Galicki,
 ``Sasakian Geometry'', Oxford Mathematical Monographs.

\bibitem{yaupriv} S.T.~Yau, private communications.

\bibitem{Klebanov:1998hh}
  I.~R.~Klebanov and E.~Witten,
  ``Superconformal field theory on threebranes at a Calabi-Yau  singularity,''
  Nucl.\ Phys.\  B {\bf 536}, 199 (1998)
  [arXiv:hep-th/9807080].

\bibitem{Itzhaki:1998dd}
  N.~Itzhaki, J.~M.~Maldacena, J.~Sonnenschein and S.~Yankielowicz,
  ``Supergravity and the large N limit of theories with sixteen
  Phys.\ Rev.\  D {\bf 58}, 046004 (1998)
  [arXiv:hep-th/9802042].

\bibitem{Aharony:2008ug}
  O.~Aharony, O.~Bergman, D.~L.~Jafferis and J.~Maldacena,
  ``N=6 superconformal Chern-Simons-matter theories, M2-branes and their
  JHEP {\bf 0810}, 091 (2008)
  [arXiv:0806.1218 [hep-th]].

\bibitem{Bagger:2006sk}
  J.~Bagger and N.~Lambert,
  ``Modeling multiple M2's,''
  Phys.\ Rev.\  D {\bf 75}, 045020 (2007)
  [arXiv:hep-th/0611108].

\bibitem{Gustavsson:2007vu}
  A.~Gustavsson,
  ``Algebraic structures on parallel M2-branes,''
  arXiv:0709.1260 [hep-th].

\bibitem{Bagger:2007jr}
  J.~Bagger and N.~Lambert,
  ``Gauge Symmetry and Supersymmetry of Multiple M2-Branes,''
  Phys.\ Rev.\  D {\bf 77}, 065008 (2008)
  [arXiv:0711.0955 [hep-th]].

\bibitem{Jafferis:2008qz}
  D.~L.~Jafferis and A.~Tomasiello,
  ``A simple class of N=3 gauge/gravity duals,''
  JHEP {\bf 0810}, 101 (2008)
  [arXiv:0808.0864 [hep-th]].

\bibitem{Martelli:2008rt}
  D.~Martelli and J.~Sparks,
  ``Notes on toric Sasaki-Einstein seven-manifolds and AdS$_4$/CFT$_3$,''
  JHEP {\bf 0811}, 016 (2008)
  [arXiv:0808.0904 [hep-th]].

\bibitem{Martelli:2008si}
  D.~Martelli and J.~Sparks,
  ``Moduli spaces of Chern-Simons quiver gauge theories and AdS(4)/CFT(3),''
  arXiv:0808.0912 [hep-th].

\bibitem{Hanany:2008cd}
  A.~Hanany and A.~Zaffaroni,
  ``Tilings, Chern-Simons Theories and M2 Branes,''
  JHEP {\bf 0810}, 111 (2008)
  [arXiv:0808.1244 [hep-th]].

\bibitem{Ueda:2008hx}
  K.~Ueda and M.~Yamazaki,
  ``Toric Calabi-Yau four-folds dual to Chern-Simons-matter theories,''
  JHEP {\bf 0812}, 045 (2008)
  [arXiv:0808.3768 [hep-th]].

\bibitem{Imamura:2008qs}
  Y.~Imamura and K.~Kimura,
  ``Quiver Chern-Simons theories and crystals,''
  JHEP {\bf 0810}, 114 (2008)
  [arXiv:0808.4155 [hep-th]].

\bibitem{Hanany:2008fj}
  A.~Hanany, D.~Vegh and A.~Zaffaroni,
  ``Brane Tilings and M2 Branes,''
  arXiv:0809.1440 [hep-th].

\bibitem{Gaiotto:2007qi}
  D.~Gaiotto and X.~Yin,
  ``Notes on superconformal Chern-Simons-matter theories,''
  JHEP {\bf 0708}, 056 (2007)
  [arXiv:0704.3740 [hep-th]].

\bibitem{Horowitz:2008bn}
  G.~T.~Horowitz and M.~M.~Roberts,
  ``Holographic Superconductors with Various Condensates,''
  Phys.\ Rev.\  D {\bf 78}, 126008 (2008)
  [arXiv:0810.1077 [hep-th]].

\bibitem{wen}
P.~A.~Lee, N.~Nagaosa and X.-G.~Wen,
``Doping a Mott insulator: Physics of high temperature superconductivity'',
Rev. Mod. Phys. {\bf 78}, 17 (2006).
[arXiv:cond-mat/0410445].

\end{thebibliography}
\end{document}